\documentclass[10pt,twocolumn,letterpaper]{article}
\pdfoutput=1
\usepackage{cvpr}
\usepackage{times}
\usepackage{epsfig}
\usepackage{graphicx}
\graphicspath{{figures/}{images/}}
\usepackage{amsmath}
\usepackage{amssymb}
\usepackage{verbatim,float}
\usepackage{amsthm}
\usepackage{algorithmic}
\usepackage{algorithm}

\newtheorem{dfn}{Definition}
\newtheorem{thm}[dfn]{Theorem}
\newtheorem{lem}[dfn]{Lemma}
\newcommand{\alg}{\textsc{HoCToP\,}}
\newcommand{\R}{\mathbb{R}}
\newcommand{\Z}{\mathbb{Z}}
\newcommand{\DT}{\mathrm{Del}}
\newcommand{\PD}{\mathrm{PD}}
\newcommand{\PB}{\mathrm{PB}}
\newcommand{\PS}{\mathrm{PS}}

\newcommand{\Rips}{\mathrm{Rips}}
\newcommand{\Cech}{\mathrm{\breve{C}ech}}

\newcommand{\rt}{\mathrm{root}}
\newcommand{\birth}{\mathrm{birth}}
\newcommand{\death}{\mathrm{death}}
\newcommand{\length}{\mathrm{length}}
\newcommand{\weight}{\mathrm{weight}}
\newcommand{\forest}{\mathrm{Forest}}
\newcommand{\minhfs}{\mathrm{minhfs}}
\newcommand{\maxhfs}{\mathrm{maxhfs}}
\newcommand{\rad}{\mathrm{rad}}
\newcommand{\al}{\alpha}
\newcommand{\ga}{\gamma}
\newcommand{\ep}{\varepsilon}
\usepackage[pagebackref=true,breaklinks=true,letterpaper=true,colorlinks,bookmarks=false]{hyperref}

\cvprfinalcopy 
\def\cvprPaperID{1997} 

\ifcvprfinal\fi
\begin{document}

\title{
A fast and robust algorithm to count topologically persistent holes in noisy clouds
}

\author{Vitaliy Kurlin\\ 
Durham University\\
Department of Mathematical Sciences, Durham, DH1 3LE, United Kingdom\\
{\tt\small vitaliy.kurlin@gmail.com, \url{http://kurlin.org}}
}

\maketitle
\thispagestyle{plain}
\pagestyle{plain}

\begin{abstract}
Preprocessing a 2D image often produces a noisy cloud of interest points.
We study the problem of counting holes in noisy clouds in the plane. 
The holes in a given cloud are quantified by the topological persistence of their boundary contours when the cloud is analyzed at all possible scales.
\smallskip

We design the algorithm to count holes that are most persistent in the filtration of offsets (neighborhoods) around given points.
The input is a cloud of $n$ points in the plane without any user-defined parameters.
The algorithm has $O(n\log n)$ time and $O(n)$ space.
The output is the array (number of holes, relative persistence in the filtration).
\smallskip

We prove theoretical guarantees when the algorithm finds the correct number of holes (components in the complement) of an unknown shape approximated by a cloud.  
\end{abstract}


\section{Introduction: counting holes in noisy clouds}
\label{sec:intro}

We apply methods from the new area of topological data analysis to counting persistent holes in a noisy cloud of points.
Such a cloud can be obtained by selecting interest points in a gray scale or RGB image. 
Our region-based method uses global topological properties of contours.
\smallskip

\begin{figure}[h]
\begin{center}
\includegraphics[width=1.0\linewidth]{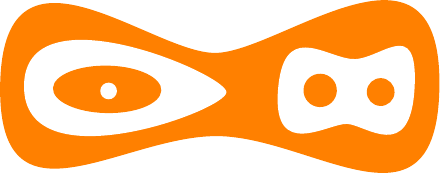}
\end{center}
\caption{The orange shape $X\subset\R^2$ with 3 white holes
 of different forms: a small disk, a ring-like hole, a `figure-eight' hole.}
\label{fig:shape-3holes}
\end{figure}

By a \emph{shape} we mean any subset $X\subset\R^2$ that can be split into finitely many (topological) triangles. 
Hence $X$ is bounded, but may not be connected.
Then a \emph{hole} in a shape $X\subset\R^2$ is a bounded connected component of the complement $\R^2-X$.
Such a hole can be a disk, a ring or may have a more complicated topological form, see Fig.~\ref{fig:shape-3holes}.
\smallskip

\begin{figure}[h]
\begin{center}
\includegraphics[width=1.0\linewidth]{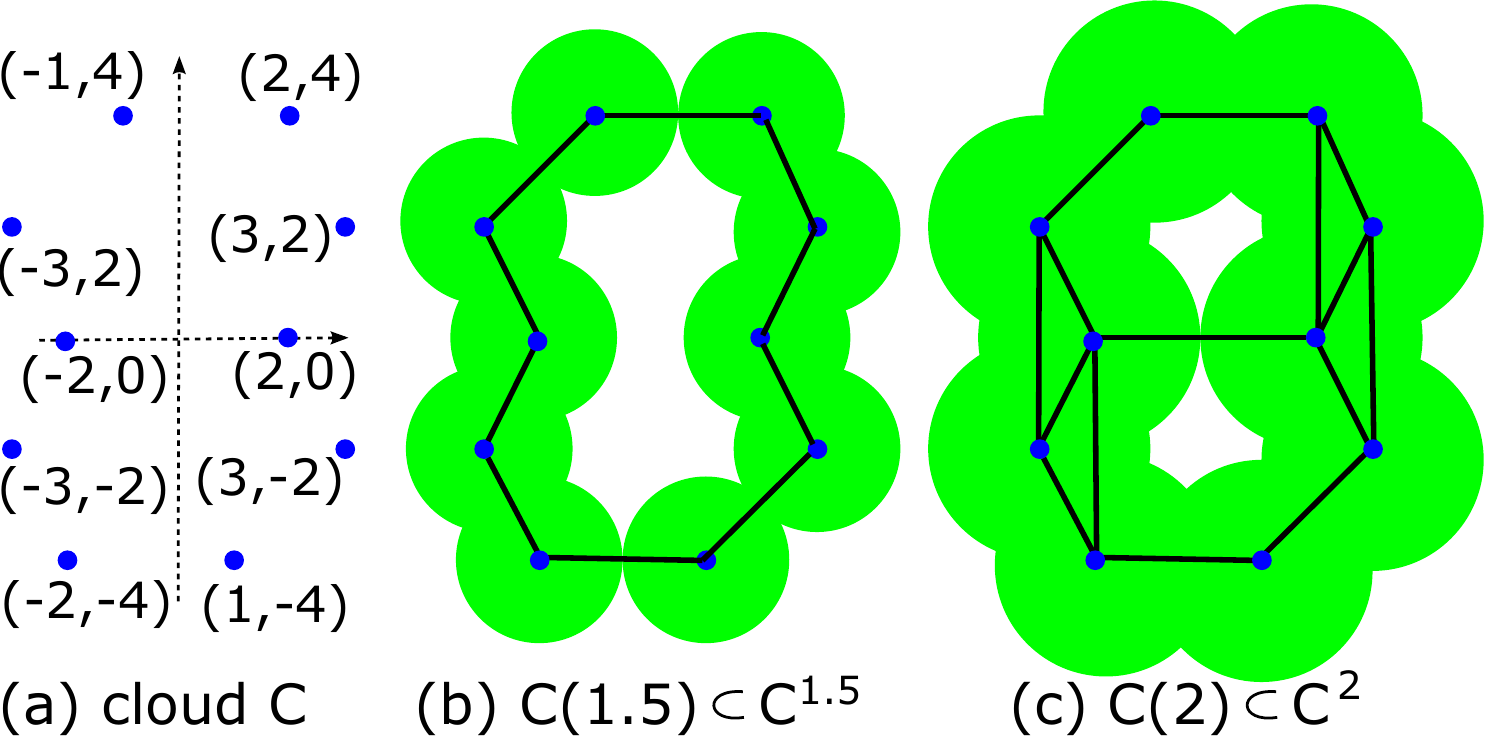}
\end{center}
\caption{The big hole in the green offset $C^{\al}$ is born at $\al=1.5$, splits into 2 smaller holes at $\al=2$ and dies at $\al\approx 2.577$, so the topological persistence of this hole is $\death-\birth\approx 1.077$.}
\label{fig:figure-eight-offsets}
\end{figure}

The \emph{$\al$-offset} $X^{\al}$ is the union $\cup_{p\in X} B(p;\al)$ of disks with the radius $\al\geq 0$ and centers at all $p\in X$.
For instance, $X^0$ is the original shape $X\subset\R^2$.
When $\al$ is increasing, the holes of $\R^2-X^{\al}$ are shrinking,
 may split into smaller newborn holes and will eventually die, each at its own death time $\al$, see Fig.~\ref{fig:figure-eight-offsets}.
The \emph{persistence} of a hole is its life span $\death-\birth$ in the filtration $\{X^{\al}\}$ of all $\al$-offsets.
So we quantify holes by their persistence at different scales $\al$.
\smallskip

\noindent
{\bf Hole counting problem.}
Let a shape $X$ be represented by a finite sample $C$ of points in $\R^2$.
Find conditions on $X$ and its sample when one can quickly count persistent holes.
\smallskip

We solve the problem by the algorithm $\alg$: Hole Counting based on Topological Persistence.
The only input is a finite cloud $C$ of $n$ points approximating an unknown shape $X\subset\R^2$.
The algorithm outputs the relative persistence of $k$ holes in the filtration $\{C^{\al}\}$ for all $k\geq 0$.
If the scale $\al$ is random and uniform, this output gives probabilities $P(k\mbox{ holes})$.
The boundary edges of persistent holes can be quickly post-processed to extract all boundary contours.

\begin{figure}[h]
\begin{center}
\includegraphics[width=1.0\linewidth]{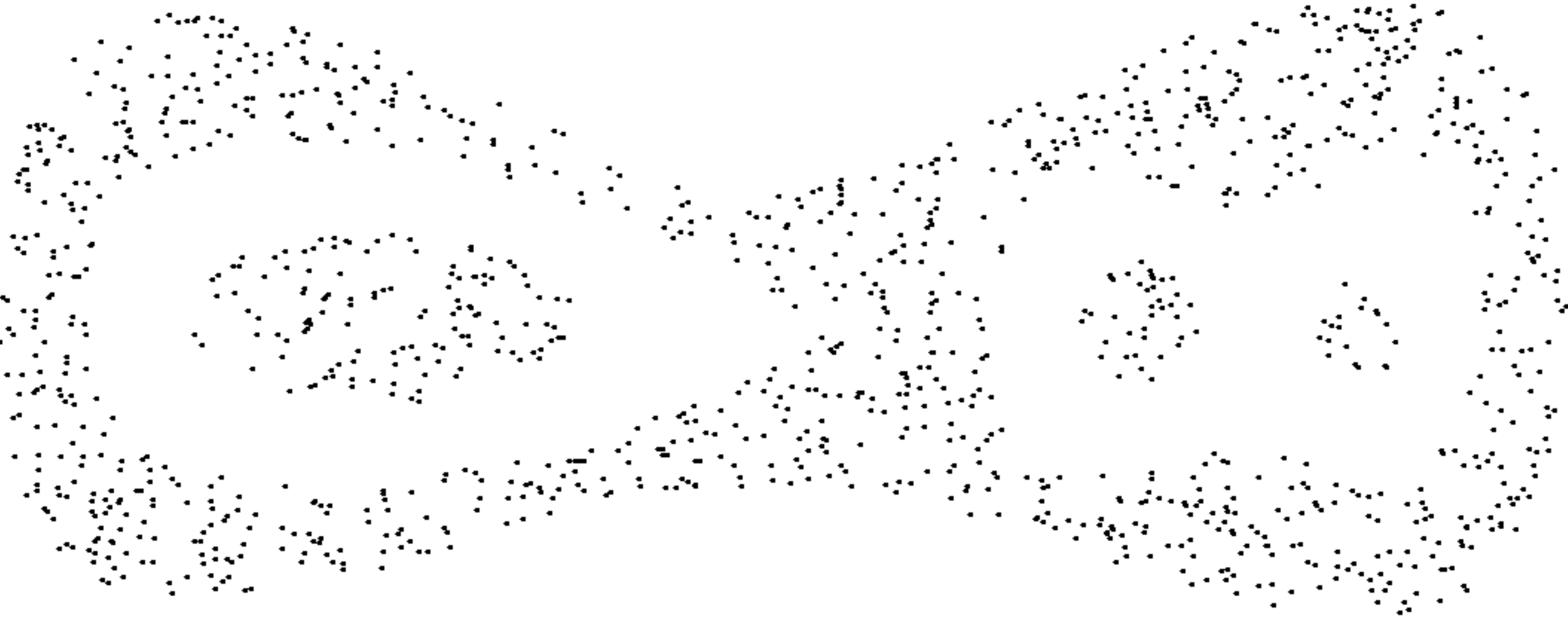}
\end{center}
\caption{Input: cloud $C$ of 1251 points uniformly sampled from the shape with 3 holes in Fig.~\ref{fig:shape-3holes}. Output probabilities of \alg: 
 $P(\mbox{3 holes})\approx 24\%$,
 $P(\mbox{2 holes})\approx 13\%$,
 $P(\mbox{8 holes})\approx 11\%$.}
\label{fig:1251C-yellow-shape2422}
\end{figure}

Theorems~\ref{thm:algorithm}, \ref{thm:guarantee} say that the algorithm \alg quickly and correctly finds all persistent holes using only a good enough sample $C$ of an unknown shape $X$, see section~\ref{sec:results}.

\section{Main results: the algorithm and guarantees}
\label{sec:results}

We start from a high-level description of our algorithm.
\smallskip

The topological persistence of contours in the filtration $\{C^{\al}\}$ is computed by using a Delaunay triangulation $\DT(C)$ of a given cloud $C\subset\R^2$ of $n$ points.
By Nerve Lemma~\ref{lem:nerve} the $\al$-offsets $C^{\al}$ can be continuously deformed to the $\al$-complexes $C(\al)$, which filter $\DT(C)$ as follows:\\ $C=C(0)\subset\dots\subset C(\al)\subset\dots\subset C(+\infty)=\DT(C)$.
Each $C(\al)$ has some edges and triangles from $\DT(C)$.

The graph dual to $\DT(C)$ is filtered by the subgraphs $C^*(\al)$ whose connected components correspond to holes in $C(\al)$.
When $\al$ is decreasing, $C(\al)$ is shrinking, so its holes are growing and corresponding components of $C^*(\al)$ merge at critical values of $\al$, see Fig.~\ref{fig:4points}.
The persistence of cycles in the filtration $\{C^{\al}\}$ corresponds to the persistence of components in $\{C^*(\al)\}$, see Duality Lemma~\ref{lem:duality}.
\smallskip

The pairs $(\birth,\death)$ of connected components in $\{C^*(\al)\}$ are found via a union-find structure by adding edges and merging components.
So computing the 1-dimensional persistence of cycles in $\{C^{\al}\}$ reduces to the 0-dimensional persistence of components in $\{C^*(\al)\}$.
\smallskip

Starting from a given cloud $C\subset\R^2$ of $n$ points with real coordinates $(x_i,y_i)$, $i=1,\dots,n$, we find a Delaunay triangulation $\DT(C)$ in $O(n\log n)$ time with $O(n)$ space.
Then we remove each edge of $\DT(C)$ one by one in the decreasing order of their length.
Removing an edge may break a contour when adjacent regions in $C(\al)$ and the corresponding components of $C^*(\al)$ merge.
In the case of a merger, a younger component of $C^*(\al)$ and the corresponding hole in $C(\al)$ die. 
We note the $\birth$ and $\death$ of each dead hole.
We get the probability of $k$ holes as the relative length of
 all intervals of the scale $\al$ when $C^{\al}\subset\R^2$ has $k$ holes.

\begin{thm}
\label{thm:algorithm}
The algorithm $\alg$ counts all holes in a given cloud $C\subset\R^2$ of $n$ points in $O(n\log n)$ time with $O(n)$ space.
All holes are ordered by their topological persistence in the ascending filtration $\{C^{\al}\}$ of the $\al$-offsets.
\end{thm}

\begin{dfn}[$\ep$-sample]
A cloud $C$ is an \emph{$\ep$-sample} of a shape $X\subset\R^2$ if $X\subset C^{\al}$ and $C\subset X^{\al}$.
So any point of $C$ is within the distance $\ep$ from a point of $X$ and any point of $X$ is at most $\ep$ away from a point of $C$.
Hence $\ep$ can be considered as the upper bound of some arbitrary noise.
\end{dfn}

\begin{dfn}[min and max homological feature sizes]
For any shape $X\subset\R^2$, let $\al=\minhfs(X)$ be the minimum \emph{homological feature size} when a first hole is born or dies in $X^{\al}$.
Let $\al=\maxhfs(X)$ be the maximum \emph{homological feature size}
 after which no holes are born or die in $X^{\al}$.
\end{dfn}

Theorem~\ref{thm:guarantee} gives sufficient (not necessary) conditions when the algorithm finds the correct number of holes in an unknown shape $X\subset\R^2$ that is represented by its finite sample $C$.  
We extend the Homology Inference Theorem~\cite{CSEH07} to the case when the upper bound $\ep$ of noise is unknown.

\begin{thm}
\label{thm:guarantee}
Let a cloud $C$ be an $\ep$-sample of a shape $X\subset\R^2$ 
 with an unknown parameter $\ep$ such that $\minhfs(X)>\frac{1}{2}\maxhfs(X)+4\ep$.
If no new holes are appear in $X^{\al}$ when $\al$ is increasing, then the algorithm $\alg$ finds the correct number of holes in $X$ by using only the cloud $C$.
\end{thm}

The condition $\minhfs(X)>\frac{1}{2}\maxhfs(X)+4\ep$ means that all holes of $X$, which are bounded components of $\R^2-X$, have comparable sizes (neither tiny nor huge).
\smallskip

Even if the conditions of Theorem~\ref{thm:guarantee} are not satisfied, we can always find the number $k$ of holes with the highest probability.
The algorithm $\alg$ can also accept a signal-to-noise ratio $\tau$ and output all holes whose persistence is larger than $\tau$.
Alternatively, the user may prefer to get most likely outputs ordered by the probability $P(k\mbox{ holes})$.

\section{Previous work on computing persistence}
\label{sec:previous}

The offsets $C^{\al}$ of a finite cloud $C$ are usually studied through the \u{C}ech or Rips complexes, which may contain up to $O(n^k)$ simplices in all dimensions $k\leq n-1$ even if $C\subset\R^2$.
A Delaunay triangulation has the advantage of a smaller size up to $m=O(n^2)$ in dimensions $n=2,3,4$.
\smallskip

The fastest algorithm \cite{MMS11} for computing persistence of a filtration in all dimensions has the same running time $O(m^{2.376})$ in the number $m$ of simplices as the best known time for the multiplication of two $m\times m$ matrices.
\smallskip

In dimension 0 the persistence can be computed in almost linear time \cite[p.~6--8]{EH10}, 
 which was used for simplifying functions on surfaces \cite{AGHLM09} and for 
 approximating persistence of an unknown scalar field from its values on a sample \cite{CGOS11}. 
\smallskip

Two extra parameters were used in a Delaunay-based image segmentation \cite{LF07}: $\al$ for the radius of disks centered at  points of a cloud $C$ and $p$ for a desired level of persistence.

\section{Delaunay triangulation and $\al$-complexes}
\label{sec:complexes}


\begin{dfn}[simplicial complex]
\label{dfn:2dim-complex}
A \emph{simplicial 2-complex} is a finite set of \emph{simplices} (vertices, edges, triangles):
\smallskip

\noindent
$\bullet$
the sides of any triangle are included in the complex;
\smallskip

\noindent
$\bullet$
the endpoints of any edge are included in the complex;
\smallskip

\noindent
$\bullet$
two triangles can intersect only along a common edge;
\smallskip

\noindent
$\bullet$
edges can meet only at a common endpoint (a vertex);
\smallskip

\noindent
$\bullet$
an edge can not pierce through the interior of a triangle.
\end{dfn}

If a complex $S$ is drawn in $\R^n$ without self-intersections, we may call this image $|S|$ a \emph{geometric realization} of $S$.
We have defined a shape $X\subset\R^2$ as a geometric realization
 of a 2-complex.
For instance, a round disk whose boundary is split into 3 edges by 3 vertices is a topological triangle.
\smallskip

A \emph{cycle} in a complex is a sequence of edges $e_1,\dots,e_m$ such that any consecutive edges $e_i,e_{i+1}$ (in the cyclic order) have a common vertex.
Any loop in a geometric realization $|S|$ continuously deforms to a cycle of edges in $S$. 

\begin{dfn}[Delaunay triangulation $\DT$]
\label{dfn:Delaunay}
For a point $p_i$ in a cloud $C=\{p_1,\dots,p_n\}\subset\R^2$,
 the \emph{Voronoi cell} 
$V(p_i)=\{q\in\R^2 : d(p_i,q)\leq d(p_j,q) \;\forall j\neq i\}$
 is the set of all points $q$ that are (non-strictly) closer to $p_i$ than to other points of $C$.
The \emph{Delaunay triangulation} $\DT(C)$ is the nerve of the Voronoi diagram $\cup_{p\in C} V(p)$.
Namely, $p,q,r\in C$ span a triangle if and only if $V(p)\cap V(q)\cap V(r)\neq\emptyset$.
\end{dfn}

By another definition \cite[section~9.1]{BCKO08} the circumcircle of any \emph{Delaunay triangle} in $\DT(C)$ encloses no points of $C$.
\smallskip

For a cloud $C\subset\R^2$ of $n$ points, let $\DT(C)$ have $k$ triangles and $b$ boundary edges in the external region.
Counting all $E$ edges over triangles, we get $3k+b=2E$.
Euler's formula $n-E+(k+1)=2$ implies that 
 $k=2n-b-2$, $E=3n-b-3$. 
So $\DT(C)$ has $O(n)$ edges and triangles.

\begin{dfn}[$\al$-complex $C(\al)$]
\label{dfn:alpha-complex}
For a scale parameter $\al>0$, the \emph{$\al$-complex} $C(\al)$ is the nerve of $\cup_{p\in C}(V(p)\cap B(p;\al))$, see \cite[section III.4]{EH10}. 
Points $p,q\in C$ are connected by an edge if $V(p)\cap B(p;\al)$ meets $V(q)\cap B(q;\al)$.
Three points $p,q,r\in C$ span a triangle if the intersection
  $V(p)\cap B(p;\al)\cap V(q)\cap B(q;\al)\cap V(r)\cap B(r;\al)\neq\emptyset$. 
\end{dfn}

If $\al>0$ is very small, all points of $C$ are disjoint in $C(\al)$,
 while $C(\al)=\DT(C)$ for any large enough $\al$, see  examples in Fig.~\ref{fig:figure-eight-offsets}.
So all $\al$-complexes form the filtration 
 $C=C(0)\subset\dots\subset C(\al)\subset\dots\subset C(+\infty)=\DT(C)$. 
Edges or triangles are added only at \emph{critical values} of $\al$.

\begin{lem}[Nerve of a ball covering \cite{Edel95}]
\label{lem:nerve}
The union of balls $C^{\al}=\cup_{p\in C} B(p;\al)$ continuously deforms to (has the homotopy type of) a geometric realization of $C(\al)$.
\end{lem}

\section{Persistent homology: definitions, examples}
\label{sec:persistence}

\begin{dfn}[1-dimensional homology $H_1$]
\label{dfn:1dim-homology}
We consider the 1-dimensional homology group $H_1(S)$ only with coefficients in $\Z/2\Z=\{0,1\}$.
Cycles of a 2-dimensional complex $S$ can be algebraically written as linear combinations of edges (with coefficients $0$ or $1$) and generate the vector space $C_1$ of cycles.
The boundaries of all triangles in $S$ (as cycles of 3 edges) generate the subspace $B_1\subset C_1$.
The quotient group $C_1/B_1$ is the \emph{homology} group $H_1(S)$.
\end{dfn}

By a \emph{filtration} $\{S(\al)\}$ we mean a sequence of nested complexes 
 $S(0)\subset\dots\subset S(\al)\subset\dots$ that change only at finitely many
 \emph{critical values} $\al_1,\dots,\al_m$.
Then we get the induced linear maps $H_1(S(\al_1))\to\dots\to H_1(S(\al_m))$.

\begin{dfn}[persistence diagram $\PD\{S(\al)\}$]
\label{dfn:pers-diagram}
In a filtration $\{S(\al)\}$ a homology class $\ga\in H_1(S(\al_i))$ is \emph{born} at $\al_i=\birth(\ga)$ if $\ga$ is not in the image of $H_1(S(\al))\to H_1(S(\al_i))$ 
for any $\al<\al_i$.
The class $\ga$ \emph{dies} at the first time $\al_j=\death(\ga)\geq\al_i$ when the image of $\ga$ under $H_1(S(\al_i))\to H_1(S(\al_j))$ merges into the image of $H_1(S(\al))\to H_1(S(\al_j))$ for some $\al<\al_i$. 
The class $\ga$ has the \emph{persistence} $\death(\ga)-\birth(\ga)$.
The point $(\al_i,\al_j)$ has the multiplicity $\mu_{ij}$ equal to the number of   independent classes that are born at $\al_i$ and die at $\al_j$.
The \emph{persistence diagram} $\PD\{S(\al)\}$ in $\{(x,y)\in\R^2 : x\leq y\}$ is the multi-set consisting of all points $(\al_i,\al_j)$ with the multiplicity $\mu_{ij}$
 and all diagonal points $(x,x)$ with the infinite multiplicity.
\end{dfn}

Pairs with a low persistence $\death-\birth$ (close to the diagonal $\{x=y\}$ in $\PD$) are treated as noise. 
Pairs with a high persistence represent persistent cycles in $\{S(\al)\}$.
\smallskip

\begin{figure}[h]
\begin{center}
\includegraphics[width=1.0\linewidth]{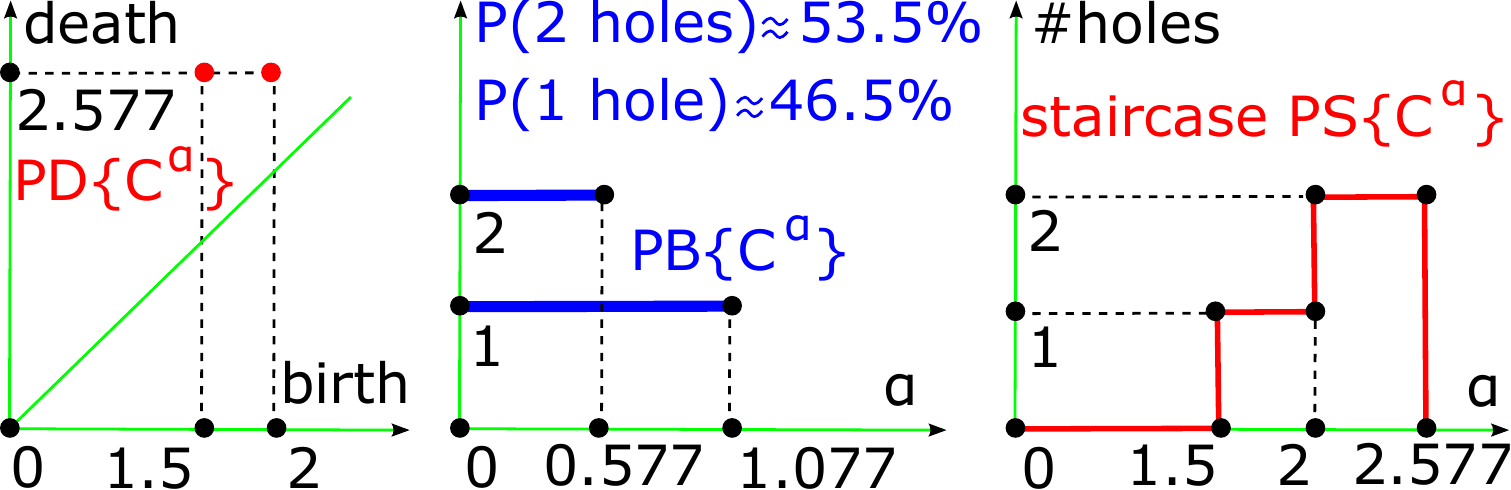}
\end{center}
\caption{Extra outputs for the cloud $C$ of 10 points in Fig.~\ref{fig:figure-eight-offsets}. Left: persistence diagram, middle: barcode, right: persistence staircase.}
\label{fig:figure-eight-persistence}
\end{figure}

\begin{figure*}[t]
\begin{center}
\includegraphics[width=0.3\linewidth]{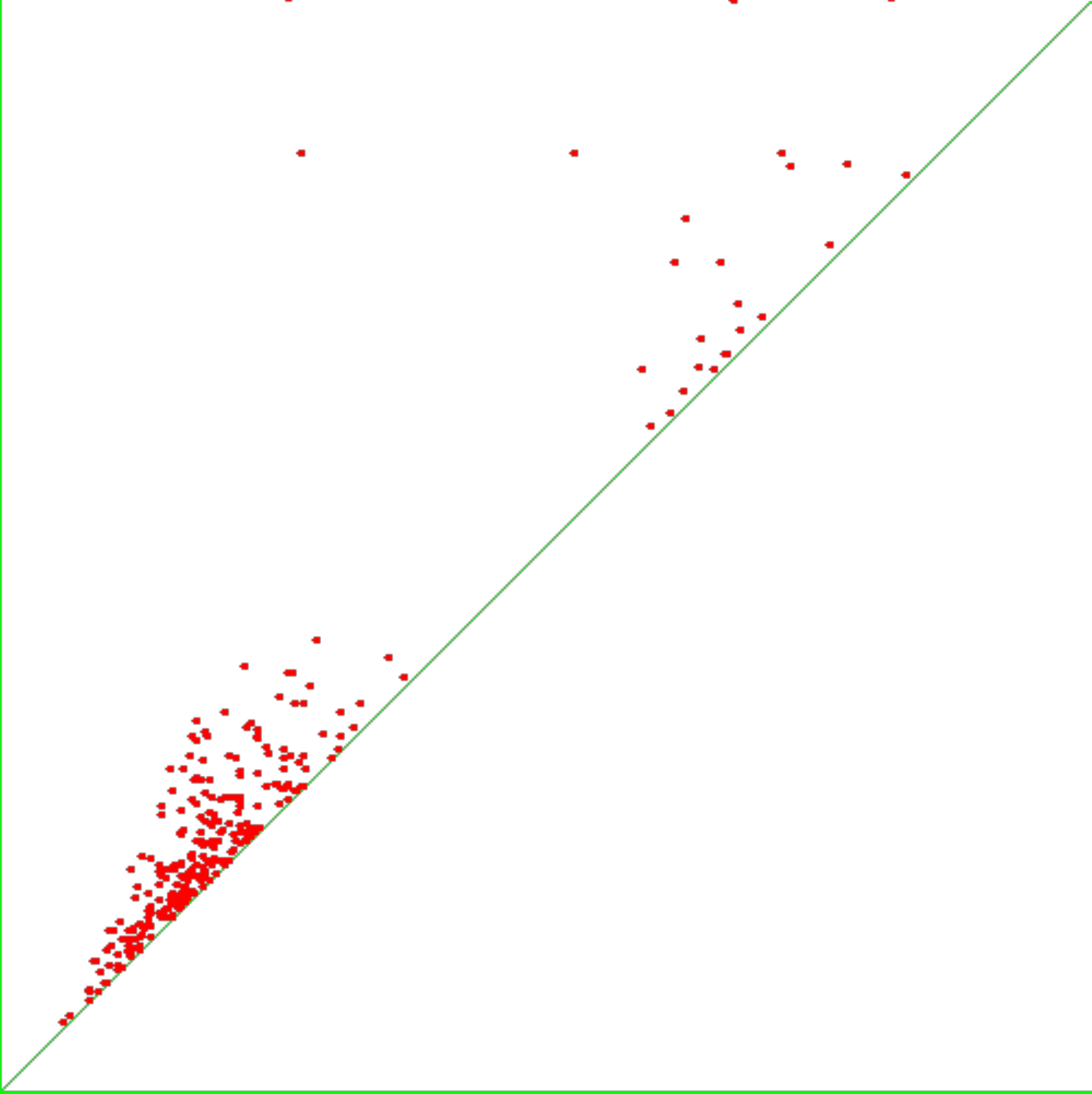}
\hspace*{0.02\linewidth}
\includegraphics[width=0.3\linewidth]{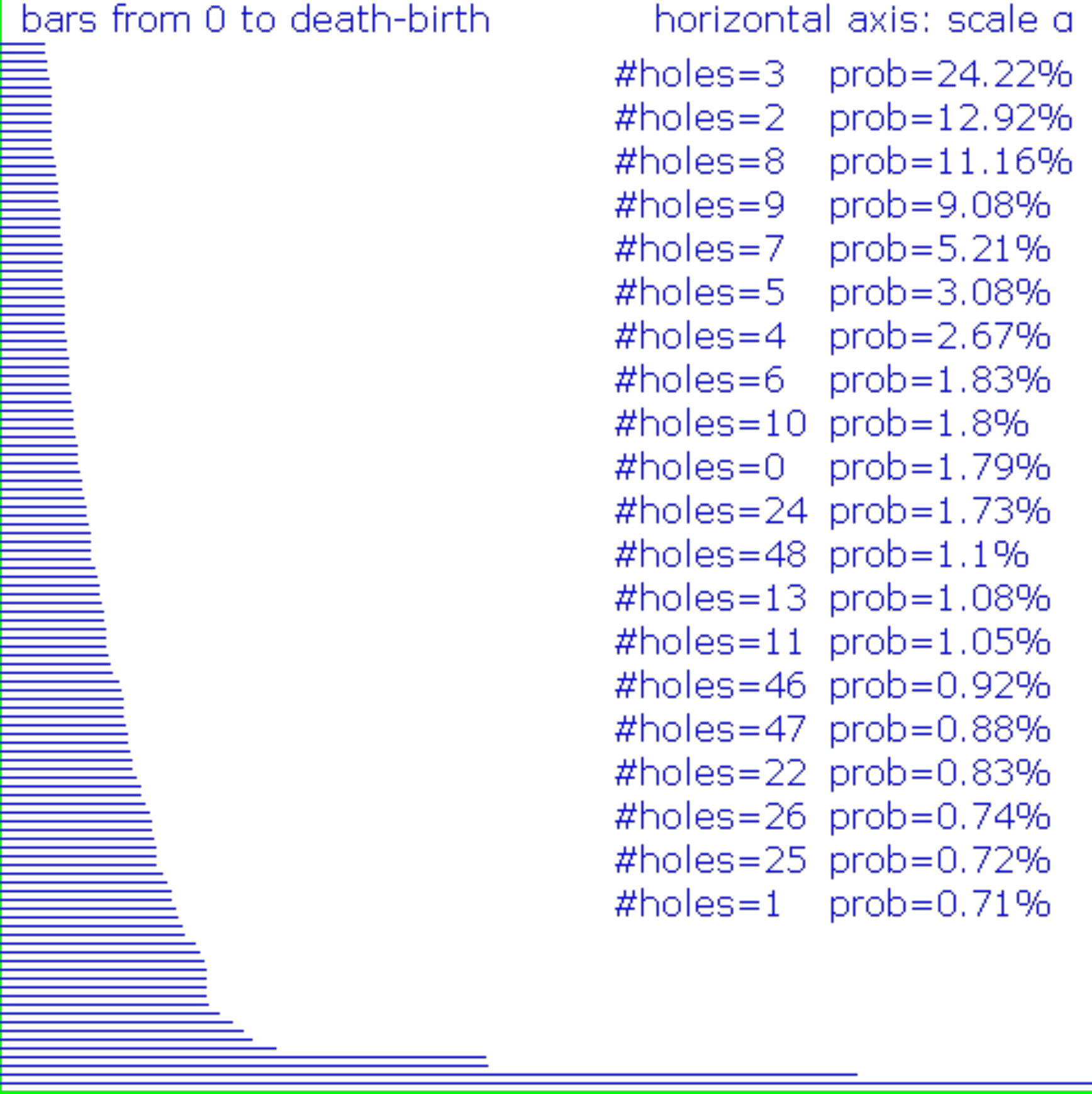}
\hspace*{0.02\linewidth}
\includegraphics[width=0.3\linewidth]{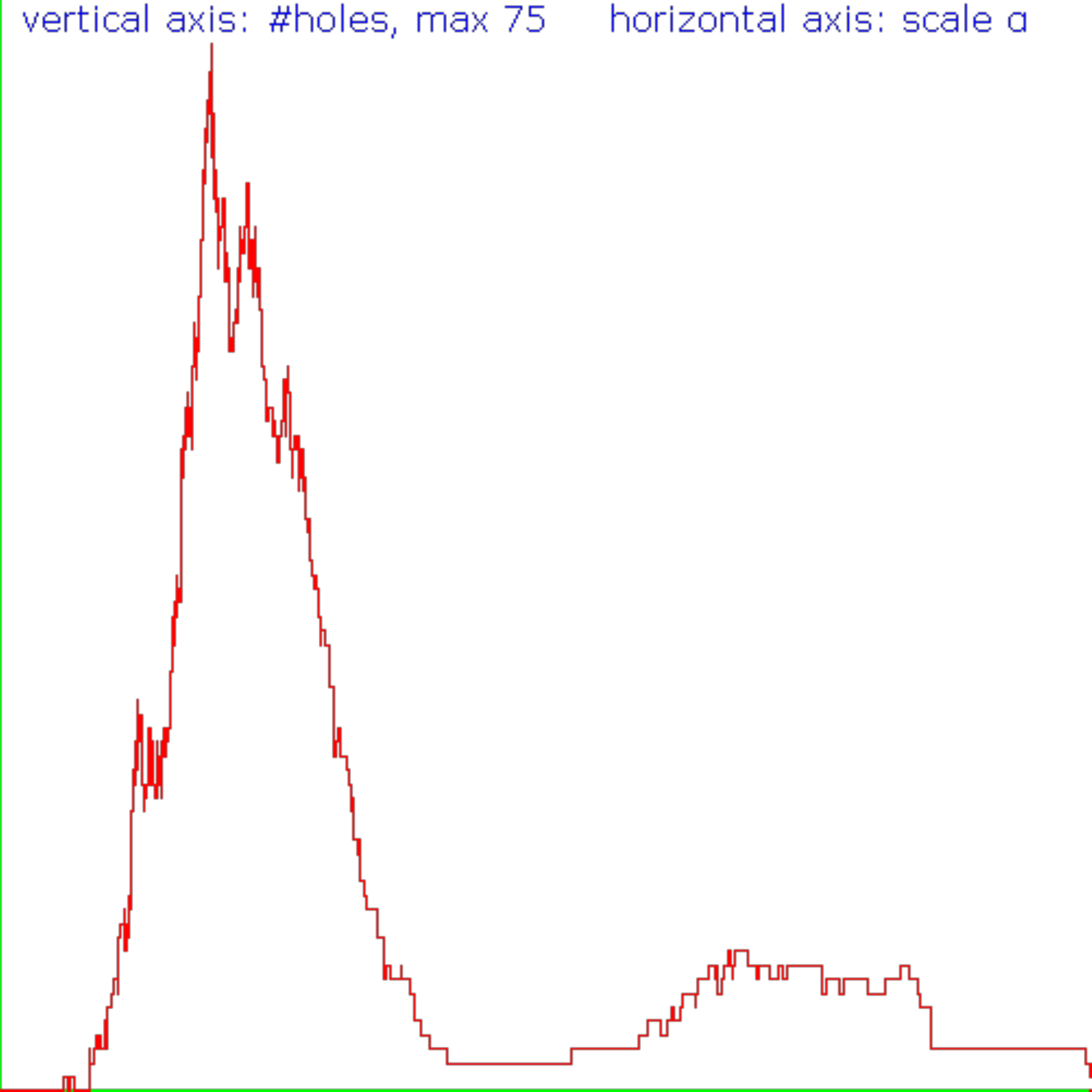}
\end{center}
\caption{Extra outputs for the cloud $C$ of 1251 points in Fig.~\ref{fig:1251C-yellow-shape2422}.
Left: persistence diagram $\PD$, middle: barcode $\PB$, right: staircase $\PS$.}
\label{fig:1251-yellow-shape2422}
\end{figure*}

We shall consider the filtrations of $\al$-offsets $\{X^{\al}\}$ and $\{C^{\al}\}$ for a shape $X\subset\R^2$ and a finite cloud $C\subset\R^2$.
Figures~\ref{fig:figure-eight-persistence} and \ref{fig:1251-yellow-shape2422} show the persistence diagram $\PD$ for the filtration of the $\al$-offsets $C^{\al}$ equivalent to $C(\al)$ by Lemma~\ref{lem:nerve}.

We can convert the persistence diagram into the persistence barcode $\PB\{C^{\al}\}$.
All pairs $(\birth,\death)$ give horizontal bars ordered by their length $\death-\birth$.
Usually the bars are drawn from the left endpoint $0$ to the right endpoint $\death-\birth$, see the middle picture in Fig.~\ref{fig:figure-eight-persistence}.
\smallskip

We suggest one more way to visualize persistence.
Each pair $(\birth,\death)$ defines the function $f(\al)=1$ for $\birth\leq\al<\death$ and $f(\al)=0$ otherwise.
The sum of these functions over all pairs gives the \emph{persistence staircase} $\PS\{C^{\al}\}$.
The value of this piecewise constant function of $\al$ is the number of holes in the offset $C^{\al}$.
We have connected consecutive horizontal segments of $\PS\{C^{\al}\}$ to get a `continuous' staircase as in the right picture of Fig.~\ref{fig:figure-eight-persistence}.
\smallskip

For the cloud $C$ of 10 points in Fig.~\ref{fig:figure-eight-offsets},
 the full range of the scale $\al$ is from the smallest critical value $\al=1.5$ (when a first hole is born) to the largest critical value $\al=\frac{5}{8}\sqrt{17}\approx 2.577$ (when both final holes die).
The output probability $P(\mbox{1 hole})\approx 46.5\%$ is the contribution of the interval $(1.5,2)$ to the full range $1.5\leq\al\leq\frac{5}{8}\sqrt{17}$.
The largest probability $P(\mbox{2 holes})\approx 53.5\%$ is the contribution of the interval $(2,\frac{5}{8}\sqrt{17})$ when $C^{\al}$ has exactly 2 holes. 
\smallskip

For the cloud $C$ of 1251 points in Fig.~\ref{fig:1251C-yellow-shape2422}, we scaled $\PB\{C^{\al}\}$ and $\PS\{C^{\al}\}$
 along the horizontal $\al$-axis and kept only the longest bars in the barcode $\PB\{C^{\al}\}$ in Fig.~\ref{fig:1251-yellow-shape2422}. 

\section{Persistent homology: stability and duality}
\label{sec:duality}

\begin{dfn}[bottleneck distance $d_B$]
Let the distance between $p=(x_1,y_1)$, $q=(x_2,y_2)$ in $\R^2$ be $||p-q||_{\infty}=\max\{|x_1-x_2|,|y_1-y_2|\}$.
The \emph{bottleneck distance} is 
 $d_B(D,D')=\inf_{\varphi}\sup_{p\in D} ||p-\varphi(p)||_{\infty}$ 
 over all bijections $\varphi:D\to D'$ between persistence diagrams $D,D'$.  
\end{dfn}


\begin{thm} \cite{CSEH07}
\label{thm:stability}
If a finite cloud $C$ of points is an $\ep$-sample of a shape $X\subset\R^2$, then $d_B(\PD\{X^{\al}\},\PD\{C^{\al}\})\leq\ep$.
\end{thm}

Stability Theorem~\ref{thm:stability} implies for barcodes $\PB$ that the endpoints of all bars are perturbed by at most $\ep$.
So a long bar can become only a bit shorter after adding noise.
\smallskip


To every triangle in the Delaunay triangulation $\DT(C)$, 
 let us associate a single abstract vertex $v_i$, $i=1,\dots,k$.
It will be convenient to call the external region of $\DT(C)$ 
 also a `triangle' and represent it by an extra vertex $v_0$.

\begin{figure}[h]
\begin{center}
\includegraphics[width=1.0\linewidth]{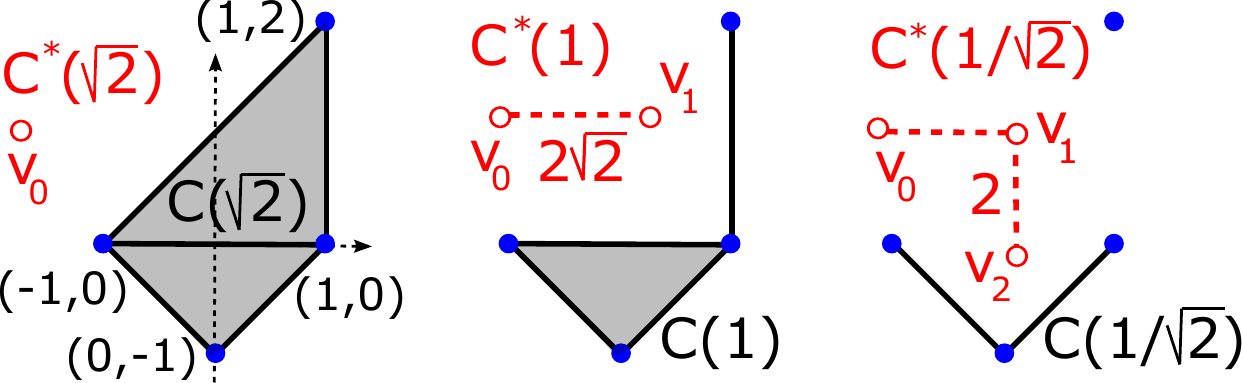}
\end{center}
\caption{The complexes $C(\al)$ have solid edges and gray triangles.
The graphs $C^*(\al)$ have circled vertices and red dashed edges.}
\label{fig:4points}
\end{figure}

\begin{dfn}[graphs $C^*(\al)$]
\label{dfn:alpha-graphs}
For any vertices $v_i, v_j$ representing adjacent triangles in $\DT(C)$,
 let $d_{ij}$ be the length of the (longest) common edge of the triangles. 
The metric graph $C^*$ dual to $\DT(C)$ has the vertices
 $v_0,v_1,\dots,v_k$ and edges of the length $d_{ij}$ connecting
 vertices $v_i,v_j$ that represent adjacent triangles,
 see Fig.~\ref{fig:4points}.
The graph $C^*$ is filtered by the subgraphs $C^*(\al)$ 
 that have only the edges of a length $d_{ij}>2\al$.
We remove any isolated node $v$ (except $v_0$) from $C^*(\al)$ if 
 the corresponding triangle $T_v$ is not acute or has 
 a small \emph{circumradius} $\rad(v)\leq\al$.
We get the filtration $C^*=C^*(0)\supset\dots\supset C^*(\al)\supset\dots\supset C^*(+\infty)=\{v_0\}$.
\end{dfn}


Components of $C^*(\al)$ are called \emph{white},
 because they represent regions in $\R^2-C(\al)$ (or holes in $\R^2-C^{\al}$).
A cycle $\ga\subset C(\al)$ is called a \emph{contour} if $\ga$ bounds a region in $\R^2-C(\al)$, so $\ga$ `encloses' the corresponding white component of $C^*(\al)$.
Lemma~\ref{lem:duality} is an analogue of the Symmetry Theorem \cite[p.~164]{EH10} for a function on a closed manifold.

\begin{lem}[Duality]
\label{lem:duality}
All contours of the complex $C({\al})$ are in 
 a 1-1 correspondence with all connected components of the graph $C^*(\al)$
 not containing the vertex $v_0$.
When $\al$ is decreasing, the contours of $C(\al)$ and the
 white components of $C^*(\al)$ have the corresponding critical moments:
\smallskip

\noindent
$\bullet$
a birth of a contour $\leftrightarrow$ a birth of a white component,
\smallskip

\noindent
$\bullet$
a death of a contour $\leftrightarrow$ a death of a white component.
\qed
\end{lem}

\section{The algorithm \alg for counting holes}
\label{sec:algorithm}


We build the union-find structure $\forest(\al)$ on the vertices of 
 the graph $C^*(\al)$.
All nodes and trees of  $\forest(\al)$ will be in a 1-1 correspondence with all vertices and white components of $C^*(\al)$.
Every node $v$ in $\forest(\al)$ has 
\smallskip

\noindent
$\bullet$
a pointer to a unique parent of the node $v$ in $\forest(\al)$;
\smallskip

\noindent
$\bullet$
a pointer to the Delaunay triangle dual to the node $v$;
\smallskip

\noindent
$\bullet$
the weight (the number of nodes below $v$ in its tree);
\smallskip

\noindent
$\bullet$
the critical value (birth) $\al_v=\sup\{\al : v\in C^*(\al)\}$.
\smallskip

If a node $v$ is a self-parent, we call $v$ a \emph{root}.
We can find $\rt(v)$ of any node $v$ by going up along parent links.
If $\al$ is decreasing, $\al_v$ can be considered as the birth time 
 when the vertex $v$ joins $C^*(\al)$.
The algorithm initializes $\forest(\al)$ as the set of isolated nodes $v_0,\dots,v_k$.
If the triangle corresponding to $v_k$ is acute, the birth time of $v_k$
 is the circumradius of the triangle, otherwise 0.
We will go through all edges of $\DT(C)$ in the decreasing order of their length and will update $\al_v$ when $v$ enters the ascending filtration $\{v_0\}=C^*(+\infty)\subset\dots\subset C^*(\al)\subset\dots\subset C^*(0)=C^*$.
\smallskip


All triangles of $C(\al)$ and the corresponding nodes 
 of $\forest(\al)$ are called \emph{gray}.
The remaining triangles and the external region of $\DT(C)$ are called \emph{white}.
The external region has birth time $+\infty$ 
 and is called a `triangle' for simplicity.
Initially all triangles with birth time 0 are gray.
\medskip

\noindent
{\bf The \emph{while loop}}.
For each edge $e\subset\DT(C)$ arriving in the decreasing order of length, 
 we find two triangles $T_u,T_v$ attached to $e$ and check if they are gray or white.
To determine if a triangle $T_v$ represented by a node $v$ is gray, 
 we go up along parent links from $v$ to $\rt(v)$.
If the birth time of $\rt(v)$ is 0, the triangle $T_v$ is still gray, otherwise white.
\smallskip

To distinguish Cases 1 and 4 below, we also check if the triangles $T_u,T_v$ 
 attached to the current edge $e$ are in the same region of $\R^2-C(\al)$.
Case~1 means that the nodes $u,v\in\forest(\al)$ belong to the same tree, 
 so $\rt(u)=\rt(v)$.
In all 4 cases the scale $\al$ goes down through the half-length 
 $\frac{1}{2}\length(e)$ of the current edge $e$ from $\DT(C)$.
\medskip

\noindent
{\bf Case 1}: $e$ has the same white region on both sides of $e$.

\noindent
$C(\al)$ loses only the open edge $e$.
The white components of $C^*(\al)$ are unchanged.
Fig.~\ref{fig:4points} illustrates Case 1 for $\al=1$ when $C(\al)$ loses the edge connecting $(1,0)$ to $(1,2)$.
\medskip

\noindent
{\bf Case 2}: the edge $e$ is in 1 gray triangle and 1 white triangle.

\noindent
Let $u,v\in C^*(\al)$ be the vertices dual to the gray triangle $T_u$ and 
 the white triangle $T_v$ attached to the current edge $e$ in $\DT(C)$.
Then the birth times are $\al_u=0$, $\al_{\rt(v)}>0$. 
\smallskip

\begin{figure}[h]
\begin{center}
\includegraphics[width=1.0\linewidth]{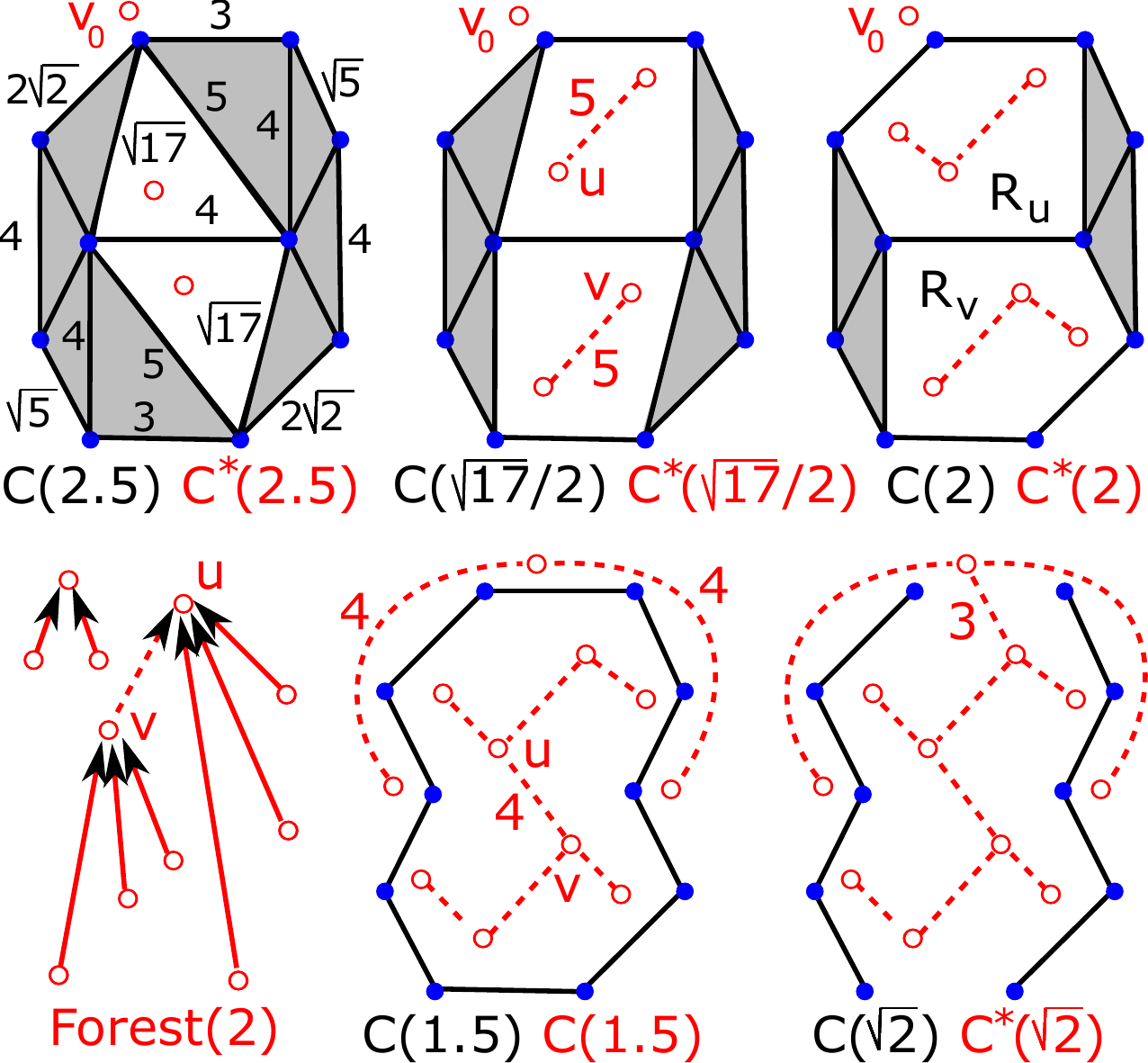}
\end{center}
\caption{Complexes $C(\al)$ and graphs $C^*(\al)$ are shown for the cloud $C$ from Fig.~\ref{fig:figure-eight-offsets}. Two trees in $\forest(\al)$ merge at $\al=2$.}
\label{fig:8filtrations}
\end{figure}

Since $\al$ is decreasing, the descending filtration $C(\al)$ 
 loses the (open) edge $e$ and the gray (open) triangle $T_u$.
So the vertex $u$ becomes connected by an edge with $v$ and joins the white component of $C^*(\al)$ containing $v$.
Then we link the isolated node $u$ to the tree containing the older node $v$ in $\forest(\al)$.
So $\rt(v)$ becomes the parent of $u$ and the weight of $\rt(v)$ jumps by 1. 
Fig.~\ref{fig:8filtrations} illustrates Case~2 for $\al=\frac{\sqrt{17}}{2}$ 
 when $C(\al)$ loses the 2 edges of length $\sqrt{17}$.

\noindent
{\bf Case 3}: the edge $e$ is in the boundary of 2 gray triangles.

\noindent
Let $u,v\in C^*(\al)$ be the vertices dual to the gray triangles $T_u,T_v$ attached to the current edge $e\subset\DT(C)$.
Then $T_u,T_v$ are right-angled triangles with the common hypotenuse $e$.
The birth time of both $u,v$ is the half-length of $e$.
Since $\al$ is decreasing, $C(\al)$ loses the (open) edge $e$ and both (open) triangles $T_u,T_v$.  
The contour $\partial(T_u\cup T_v)$ appears in $C(\al)$.
So we link the nodes $u,v$ in $\forest(\al)$.
\medskip

\noindent
{\bf Case 4}: $e$ has 2 different white regions on both sides.

\noindent
Let $u,v\in C^*(\al)$ be the vertices dual to the white triangles $T_u,T_v$ attached to the current edge $e$ in $\DT(C)$.
The descending filtration $\{C(\al)\}$ loses the (open) edge $e$.
The vertices $u,v$ become connected by an edge, so their white components in $C^*(\al)$ merge into a new big component.
By Duality Lemma~\ref{lem:duality}, two contours enclosing regions $R_u$ and $R_v$ lose their common edge $e$ and we get one larger contour $\partial(R_u\cup R_v)$ enclosing both regions.
Fig.~\ref{fig:8filtrations} illustrates Case~4 for $\al=2$ when $C(\al)$ loses the middle edge of length 4.
Then 2 white components (containing 4 vertices each) merge in the graph $C^*(\al)$
 shown after merger at $\al=1.5$.
\smallskip

To decide which white component dies, we find the roots $\rt(u),\rt(v)\in\forest(\al)$ of the trees representing $R_u,R_v$ and compare the birth times $\al_{\rt(u)},\al_{\rt(v)}$ when a first node of each tree was born.
By the elder rule \cite[p.~150]{EH10}, the older white component (say, with $u$) survives and keeps its larger birth time $\al_{\rt(u)}$.
The younger white component $R_v$ dies and we get $(\birth,\death)=(\frac{1}{2}\length(e),\al_{\rt(v)})$ for the life of the white component in the ascending filtration $\{C^*(\al)\}$ and of the corresponding contour in the descending filtration $\{C(\al)\}$.
\smallskip

We swapped the birth and death times, because the persistence is usually defined when the scale $\al$ is increasing.
However, we need the ascending filtration $\{C^*(\al)\}$ to use a union-find structure, so $\al$ is decreasing in the algorithm.
\smallskip

Finally, to merge the trees with $\rt(u),\rt(v)$ in $\forest(\al)$, we compare the weights of the roots and set the root of the (non-strictly) larger tree as the parent for the root of another tree.
So the size of any subtree grows by a factor of at least 2 each time when we pass to the parent.
We get

\begin{lem}
\label{lem:logn-paths}
By the above construction the longest path in any tree of size $k$ from $\forest(\al)$ has length $O(\log k)$.
\qed
\end{lem}

\section{Proofs of main results and our conclusion}
\label{sec:proofs}


\noindent
{\bf Proof of Theorem~\ref{thm:algorithm}}.
Constructing the Delaunay triangulation $\DT(C)$ on a cloud of $n$ points requires $O(n\log n)$ time \cite[Chapter~9]{BCKO08}.
Sorting $O(n)$ edges of $\DT(C)$ needs $O(n\log n)$ time.
Then we go through the \emph{while loop} analyzing each of the $O(n)$ edges of $\DT(C)$.
For the nodes $u,v\in\forest(\al)$ of triangles attached to each edge $e$, we find the roots of $u,v$ by going up along $O(\log n)$ parent links by Lemma~\ref{lem:logn-paths}.
All other steps in the \emph{while loop} require only $O(1)$ time.
Hence the total time is $O(n\log n)$.
The sizes of all data structures are proportional to the numbers of edges or triangles in $\DT(C)$, so we use $O(n)$ space.
\qed
\medskip

The careful analysis of a union-find structure 
says that $\forest(\al)$ can be built in time $O(nA^{-1}(n,n))$ time, where $A^{-1}(n,n)$ is the extremely slowly growing inverse Ackermann function. 
Our time $O(n\log n)$ is dominated by the construction of $\DT(C)$ and sorting all $O(n)$ edges.
\medskip

\noindent
{\bf Proof of Theorem~\ref{thm:guarantee}.}
The important critical values of $\al$ for the 1-dimensional homology
 of the filtration $\{X^{\al}\}$ are 
\smallskip


\noindent
$\bullet$
$\al=\minhfs(X)$ is the 1st value when $H_1(X^{\al})$ changes;
\smallskip

\noindent
$\bullet$
$\al=\maxhfs(X)$ is the last value when $H_1(X^{\al})$ changes.
\vspace*{-2mm}

No new holes appear in offsets $X^{\al}$ of the shape $X$ with original $k$ holes.
Then $\PD\{X^{\al}\}$ contains only points $(0,d_i)$.
The smallest death is $d_1=\minhfs(X)$.
The largest death is $d_k=\maxhfs(X)$. 
If a cloud $C$ is an $\ep$-sample of a shape $X\subset\R^2$, 
 the perturbed diagram $\PD\{C^{\al}\}$ has only points $\ep$-close to $(0,d_i)$ 
 or to the diagonal $\{x=y\}$ in the $L_{\infty}$ distance on the plane by Stability Theorem~\ref{thm:stability}.
\smallskip

The strip $\{2\ep<y-x<d_1-2\ep\}$ is the largest empty strip in $\PD\{C^{\al}\}$
due to the given condition $d_1>\frac{1}{2}d_k+4\ep$ or 
$(d_1-2\ep)-2\ep>(d_k+2\ep)-(d_1-2\ep)$.
Then we can detect this strip in $\PD\{C^{\al}\}$ without using $\ep$.  
Hence $\PD\{C^{\al}\}$ has exactly $k$ points above $y-x=d_1-2\ep$ 
 close to $(0,d_i)$ corresponding to $k$ holes of the unknown shape $X$.
\qed
\medskip


\noindent
{\bf Conclusion}.
Here are the key advantages of our approach:
\smallskip

\noindent
$\bullet$
a cloud $C\subset\R^2$ of $n$ points is simultaneously analyzed at all scales $\al$ without any extra user-defined parameters;
\smallskip

\noindent
$\bullet$
the algorithm \alg counts persistent holes of any topological form in $O(n\log n)$ time, see Theorem~\ref{thm:algorithm};
\smallskip

\noindent
$\bullet$
theoretical guarantees for a correct number of holes are proved
 for $\ep$-samples of unknown shapes, see Theorem~\ref{thm:guarantee};
\smallskip

\noindent
$\bullet$
the output is stable under perturbations of a cloud $C$ and
the only parameter of noise is an unknown upper bound $\ep$.
\medskip

\begin{figure}[h]
\begin{center}
\includegraphics[width=0.44\linewidth]{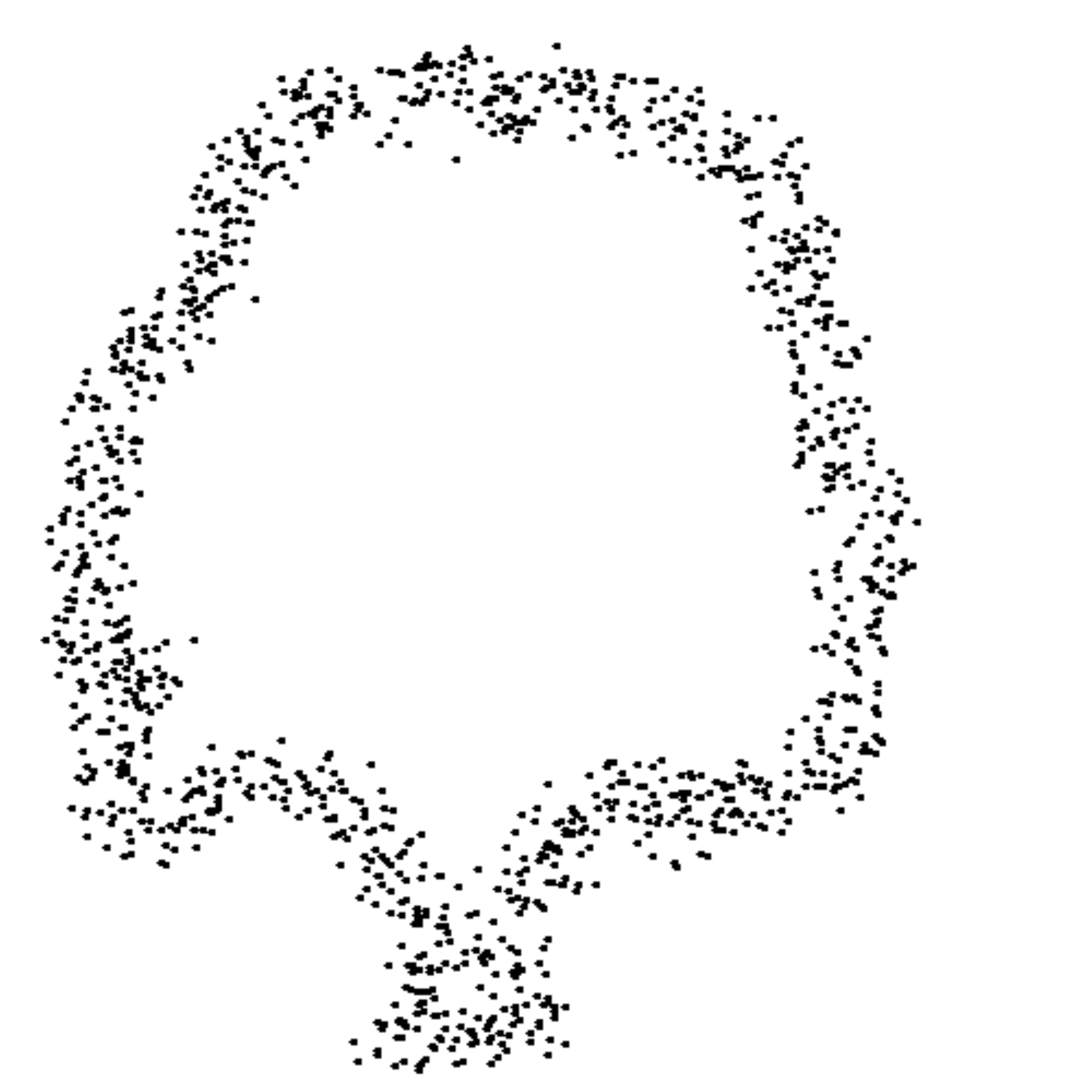}
\includegraphics[width=0.54\linewidth]{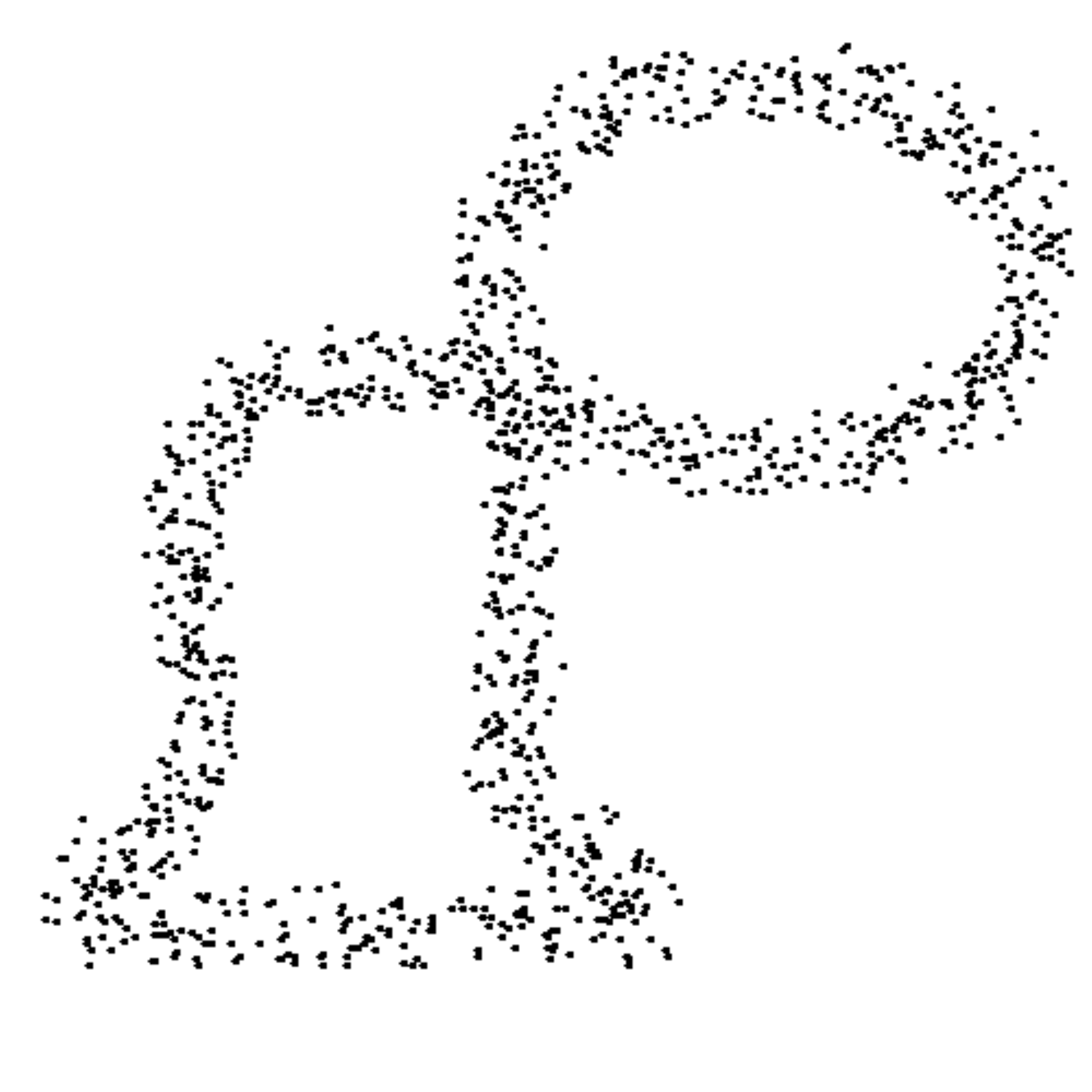}
\end{center}
\caption{Output of \alg\, for real noisy contours. 
Left: $P(\mbox{1 hole})\approx 90.5\%$, 
$P(\mbox{2 holes})\approx 3\%$, 
$P(\mbox{4 holes})\approx 0.6\%$.
Right: $P(\mbox{2 holes})\approx 74.2\%$, 
$P(\mbox{1 hole})\approx 13\%$,
$P(\mbox{3})\approx 1.3\%$.}
\label{fig:noisy-contours}
\end{figure}

Fig.~\ref{fig:noisy-contours} shows extracted contours (with our uniform noise) of 
images at \url{http://www.lems.brown.edu/~dmc}.
\vspace*{-2mm}

\noindent
More details, code, experiments are at author's website \url{http://kurlin.org}.
We thank reviewers for helpful comments and are open to collaboration on related projects.

{\small
\bibliographystyle{ieee}
\bibliography{bibdata}

\begin{thebibliography}{1}\itemsep=-1pt

\bibitem{AGHLM09}
D.~Attali, M.~Glisse, S.~Hornus, F.~Lazarus, and D.~Morozov.
\newblock Persistence-sensistive simplification of functions on surfaces in linear time.
\newblock TopoInVis 2009.

\bibitem{BCKO08}
M.~de~Berg, O.~Cheong, M.~van Kreveld, and M.~Overmars.
\newblock {\em Computational Geometry: Algorithms and Applications}.
\newblock Springer, 2008.

\bibitem{CGOS11}
F.~Chazal, L.~Guibas, S.~Oudot, P.~Skraba. 
\newblock Scalar Field Analysis over Point Cloud Data. 
\newblock Discrete and Computational Geometry, v.~46 (2011), p.743-775.

\bibitem{CSEH07}
D.~Cohen-Steiner, H.~Edelsbrunner, and J.~Harer.
\newblock Stability of persistence diagrams.
\newblock {\em Discrete and Computational Geometry}, 37:103--130, 2007.

\bibitem{Edel95}
H.~Edelsbrunner.
\newblock The union of balls and its dual shape.
\newblock {\em Discrete Computational Geometry}, 13:415--440, 1995.

\bibitem{EH10}
H.~Edelsbrunner and J.~Harer.
\newblock {\em Computational topology. An introduction}.
\newblock AMS, Providence, 2010.


\bibitem{LF07}
Letscher, D., Fritts, J.
\newblock Image segmentation using topological persistence.
\newblock Proceedings of CAIP 2007: Computer Analysis of Images and Patterns, pages 587--595.

\bibitem{MMS11}
N.~Milosavljevic, D.~Morozov, and P.~Skraba.
\newblock Zigzag persistent homology in matrix multiplication time.
\newblock Proceedings of SoCG 2011, pages 216--225, ACM.


\end{thebibliography}
}

\section*{Appendix A: a pseudo-code of \alg}

Algorithm~1 below contains the pseudo-code of the our main algorithm \alg.
Cases 2--4 from the description in section~\ref{sec:algorithm} 
 are covered in further Algorithms~2--4.

\begin{algorithm}
\label{alg:main}
\caption{Find $(\birth,\death)$ of all cycles in $C(\al)$}
\begin{algorithmic}[1]
\REQUIRE a cloud $C$ given as pairs $(x_1,y_1),\dots(x_n,y_n)$ 
\STATE Build Delaunay triangulation $\DT(C)$ with $k$ triangles
\STATE Extract all edges with pointers to 2 adjacent triangles
\STATE Sort edges of $\DT(C)$ in the decreasing order of length
\STATE $\forest\leftarrow$ isolated nodes $v_0,\dots,v_k$ with birth times $0$
\STATE For the external node $v_0$, update the birth $\al\leftarrow+\infty$
\STATE For each acute triangle $T_v$, $\al_v\leftarrow$ circumradius of $T_v$
\STATE Set the total number of links in $\forest(\al)$: $L\leftarrow 0$
\WHILE{$L<k$ (we stop when $\forest(\al)$ is a tree)}
\STATE Take the next longest edge $e$ from $\DT(C)$
\STATE Set the current critical value: $\al\leftarrow\frac{1}{2}\length(e)$
\STATE Find 2 nodes $u,v$ dual to the triangles attached to $e$
\STATE Find the roots $\rt(u),\rt(v)$ of the nodes $u,v$
\IF{$\rt(u)=\rt(v)$ ($u,v$ in the same region)}
\STATE Case 1 (no changes): continue the \emph{while loop}
\ELSIF{$\al_{\rt(u)}=0$ \AND $\al_{\rt(v)}>0$}
\STATE Case 2 ($u$ gray, $v$ white): run  Algorithm 2
\ELSIF{$\al_{\rt(u)}=0$ \AND $\al_{\rt(v)}=0$}
\STATE Case 3 (both $u,v$ are gray): run Algorithm 3
\ELSE 
\STATE Case 4 ($\al_{\rt(u)},\al_{\rt(v)}>0$): run Algorithm 4
\ENDIF
\STATE $L\leftarrow L+1$ (one link was added in Cases 2, 3, 4)
\ENDWHILE
\RETURN array of pairs $(\birth,\death)$ from Case~4
\end{algorithmic}
\end{algorithm}


Recall that a node $u$ is gray if the birth time $\al_u=0$.
The case ($u$ white, $v$ gray) is symmetric to Case~2 below,
 so we simply denote the gray node by $u$ when calling Algorithm~2.
 
\begin{algorithm}
\label{alg:case2}
\caption{Link 2 nodes $u,v$ in $\forest(\al)$ in Case 2}
\begin{algorithmic}[1]
\REQUIRE nodes $u$ and $\rt(v)$ (so $u$ is gray, $v$ is white)
\STATE Set $\rt(v)$ as parent of $u$, set $\al_u\leftarrow\al_{\rt(v)}$  
\STATE Add 1 (coming from $u$) to the weight of $\rt(v)$
\end{algorithmic}
\end{algorithm}

In Algorithm~3 below any of the gray nodes $u,v$ can be 
 the parent of the other node, we have simply chosen $u$. 

\begin{algorithm}
\label{alg:case3}
\caption{Link 2 nodes $u,v$ in $\forest(\al)$ in Case 3}
\begin{algorithmic}[1]
\REQUIRE $\al$, nodes $u,v$ dual to triangles (both $u,v$ gray)
\STATE Set $u$ as the parent of the node $v$ in $\forest(\al)$
\STATE Set: $\al_u,\al_v\leftarrow\al$, $\weight(u)\leftarrow 1$, 
 $\weight(v)\leftarrow 0$
\end{algorithmic}
\end{algorithm}

\begin{algorithm}
\label{alg:case4}
\caption{Update $\forest$ and $(\birth,\death)$ in Case 4}
\begin{algorithmic}[1]
\REQUIRE $\al$, roots $\rt(u),\rt(v)$ of white nodes $u,v$
\IF{$\al_{\rt(u)}>\al_{\rt(v)}$ (so $u$ is older than $v$)}
\STATE Add new pair $(\al,\al_{\rt(v)})$ to array $(\birth,\death)$ 
\ELSE 
\STATE Add new pair $(\al,\al_{\rt(u)})$ to array $(\birth,\death)$
\ENDIF
\IF{$\weight(\rt(u))>\weight(\rt(v))$}
\STATE $\rt(u)$ becomes the parent of $\rt(v)$ in $\forest$
\STATE Add $\weight(\rt(v))+1$ to $\weight(\rt(u))$
\ELSE 
\STATE $\rt(v)$ becomes the parent of $\rt(u)$ in $\forest$
\STATE Add $\weight(\rt(u))+1$ to $\weight(\rt(v))$
\ENDIF
\end{algorithmic}
\end{algorithm}

\section*{Appendix B: proofs of lemmas and theorems}

\noindent
{\bf Proof of Duality Lemma~\ref{lem:duality}.}
The component of $C^*(\al)$ containing the node $v_0$ 
 corresponds to the boundary contour of the external region of $\DT(C)$.
Any region of $\R^2-C(\al)$ enclosed by a contour  consists of 
 several Delaunay triangles whose dual nodes form a white component of $C^*(\al)$.
\smallskip

A birth of a contour $\ga$ in the descending filtration $\{C(\al)\}$ means that 
 $\ga$ now encloses a new region of $\R^2-C(\al)$. 
Hence a new white component is born in the dual graph $C^*(\al)$, 
 see the evolution of $C(\al),C^*(\al)$ in Fig.~\ref{fig:8filtrations}.
\smallskip

A death of a contour $\ga$ in $\{C(\al)\}$ means that $\ga$ is no longer encloses 
 a region of $\R^2-C(\al)$.
Hence two white components merge into a big one.
By the elder rule of persistence \cite[p.~150]{EH10}, the youngest component 
 dies, while the oldest component survives and inherits all nodes.
\qed
\medskip

The elder rule is a preference for the case when 
 one class has a high persistence and another has a lower persistence 
 over the case when both classes have similar persistences.
\medskip

Let us recall that Theorem~\ref{thm:algorithm} claims that the algorithm \alg runs in $O(n\log n)$ times with $O(n)$ space.
\medskip

\noindent
{\bf Step-by-step proof of Theorem~\ref{thm:algorithm}.}
Constructing the Delaunay triangulation $\DT(C)$ on a cloud of $n$ points 
 with $O(n)$ edges and triangles requires $O(n\log n)$ time and $O(n)$ space 
 \cite[Chapter~9]{BCKO08} in Steps~1--2 of Algorithm 1.
Sorting all $O(n)$ edges in the decreasing order of the length 
 needs $O(n\log n)$ time in Step 3.
Going through each of $k=O(n)$ triangles to initialize $\forest(\al)$,
 we set each birth $\al_v$ in $O(1)$ time in Steps 4--6.
Most expensive Step~12 in the \emph{while loop} is finding $\rt(u),\rt(v)$.
Each root is found recursively by going up along $O(\log n)$ parent links 
 until we come to a self-parent pointing to itself.
All other steps in Algorithms 1--4 require only $O(1)$ time.
Hence the total time of the \emph{while loop} and \alg is $O(n\log n)$. 
\qed

\section*{Appendix C: experiments on counting holes}

\begin{figure}[h]
\includegraphics[width=0.48\linewidth]{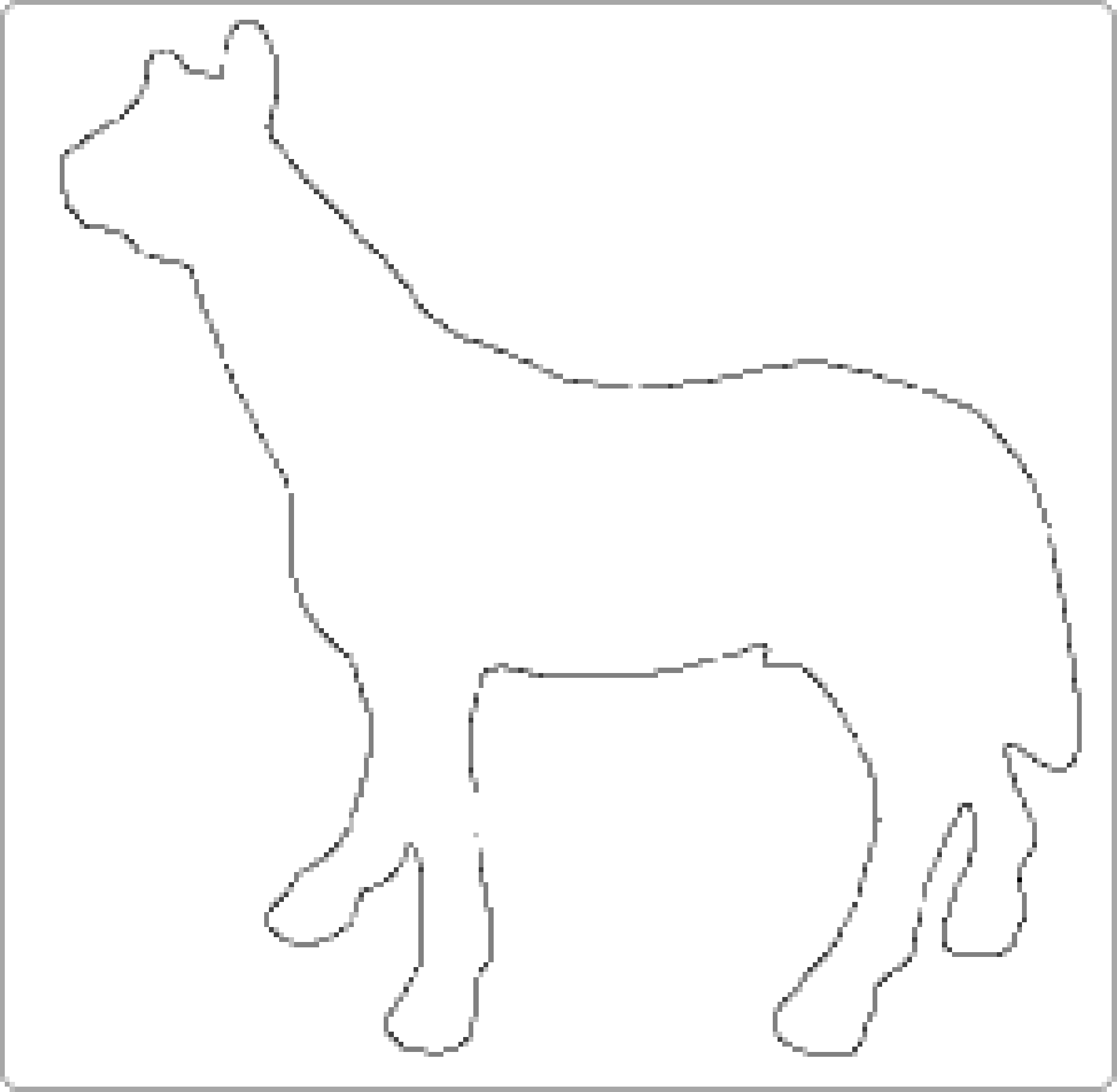}
\includegraphics[width=0.48\linewidth]{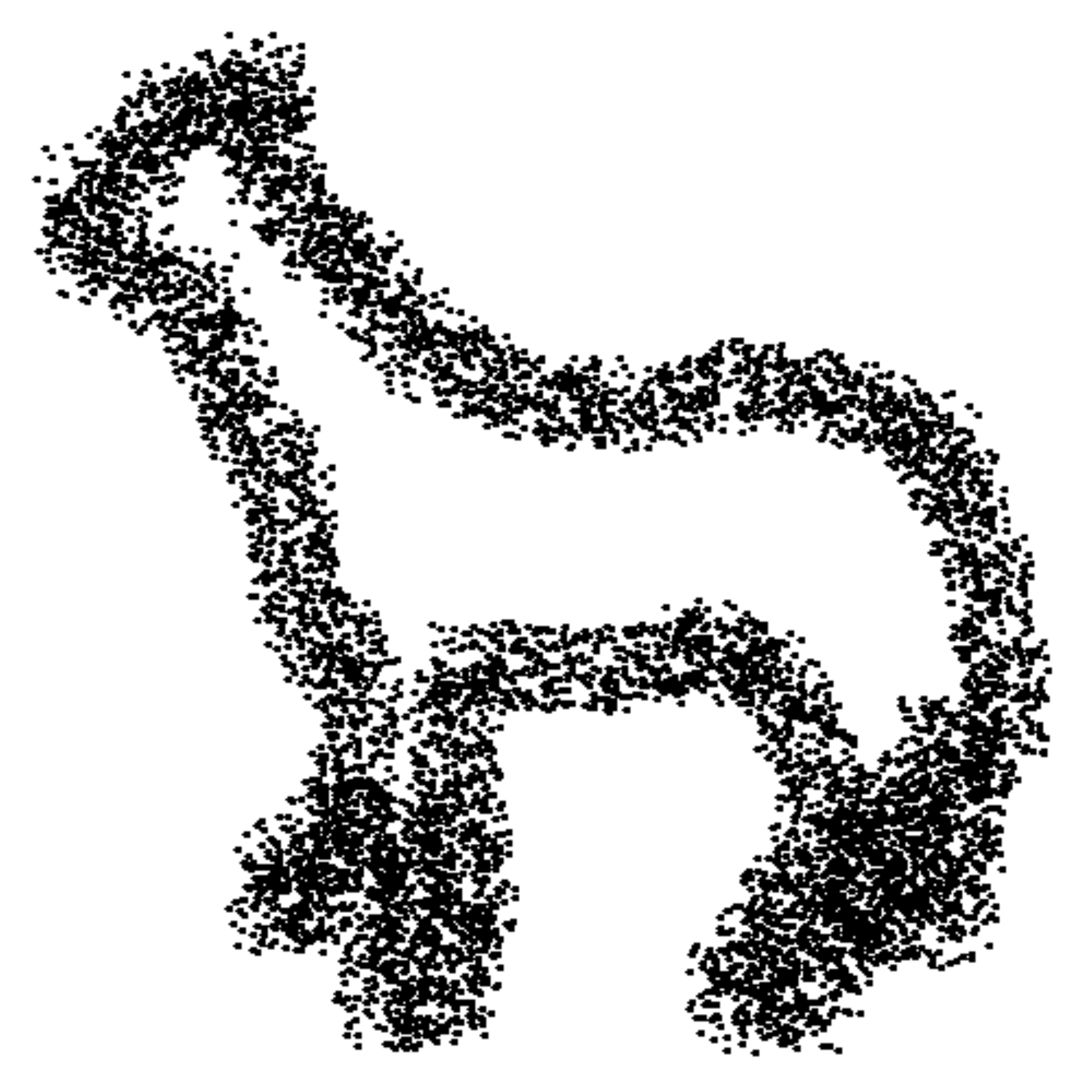}
\caption{$\quad P(\mbox{1 hole})\approx 52.7\%$, $\quad P(\mbox{2 holes})\approx 25.8\%$, 
 $P(\mbox{3 holes})\approx 9.4\%$, $\; P(\mbox{4 holes})\approx 2\%$,
 $\; P(\mbox{5 holes})\approx 0.5\%$. }
\label{fig:horse2-068-180-contour-7544n19}
\end{figure}

The left hand side picture in Fig.~\ref{fig:horse2-068-180-contour-7544n19}
 is horse2-068-180-contour.png from the database ETH80.
The right hand side picture is a cloud around the contour with added noise.
The captions contain output probabilities of \alg
 for most likely numbers of holes when the scale $\al$ is uniform.
\medskip

\begin{figure}[h]
\includegraphics[height=0.52\linewidth]{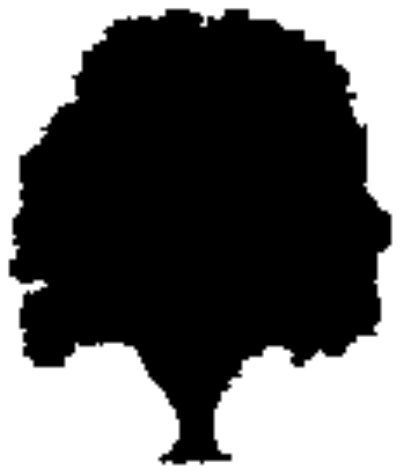}
\includegraphics[width=0.52\linewidth]{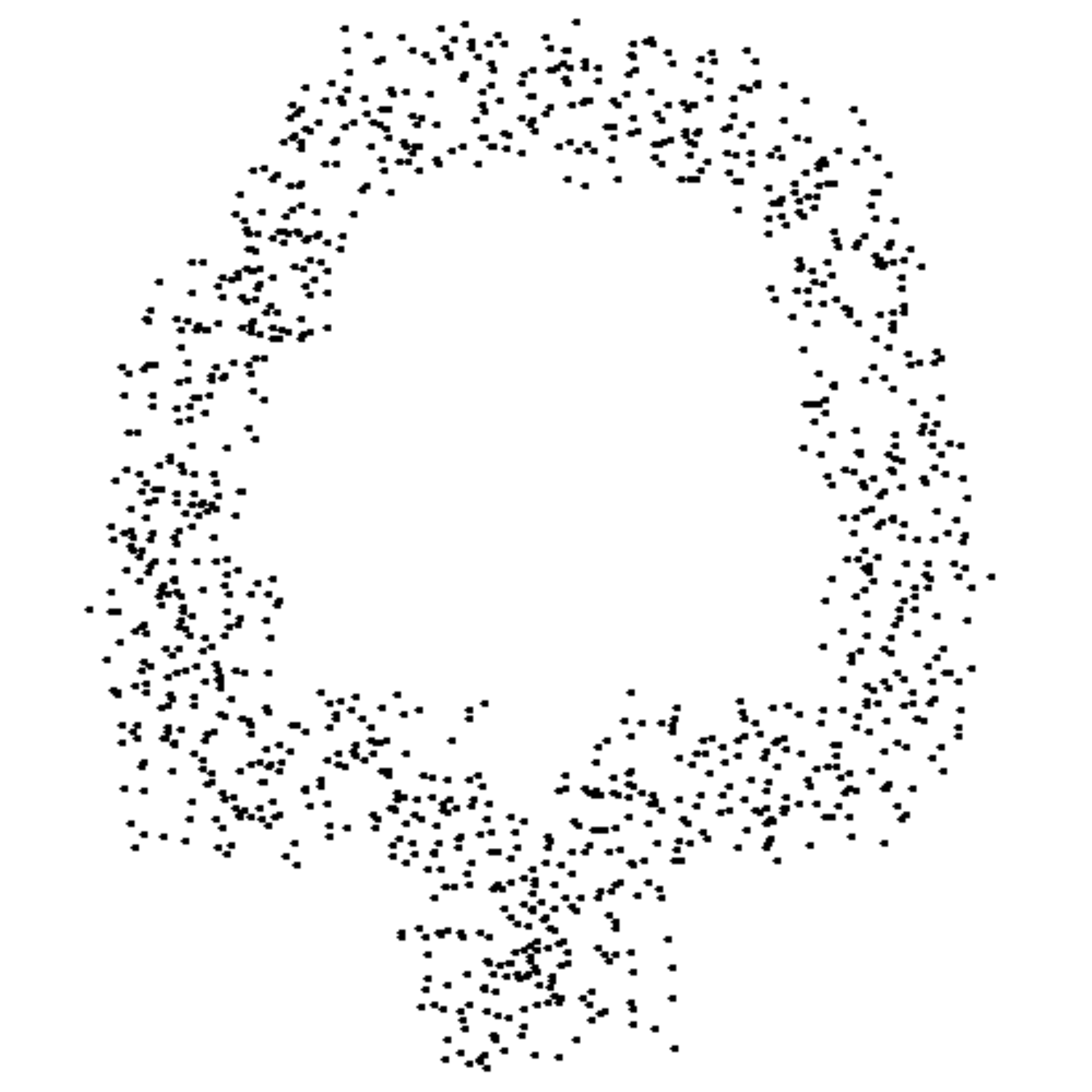}
\caption{$\quad P(\mbox{1 hole})\approx 88.4\%$, $\quad P(\mbox{2 holes})\approx 1.5\%$, 
 $P(\mbox{0 holes})\approx 0.9\%$, $\; P(\mbox{13 holes})\approx 0.5\%$,
 $\; P(\mbox{5 holes})\approx 0.4\%$. }
\label{fig:Tree004-1217n10}
\end{figure}

\begin{figure}[h]
\includegraphics[height=0.6\linewidth]{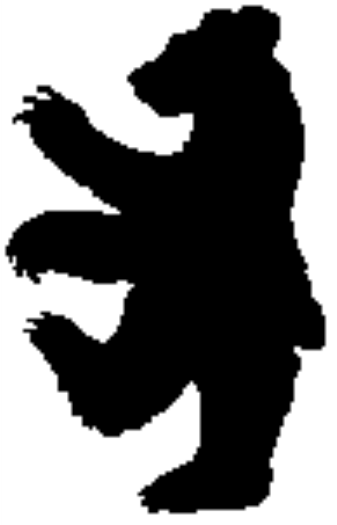}
\includegraphics[height=0.6\linewidth]{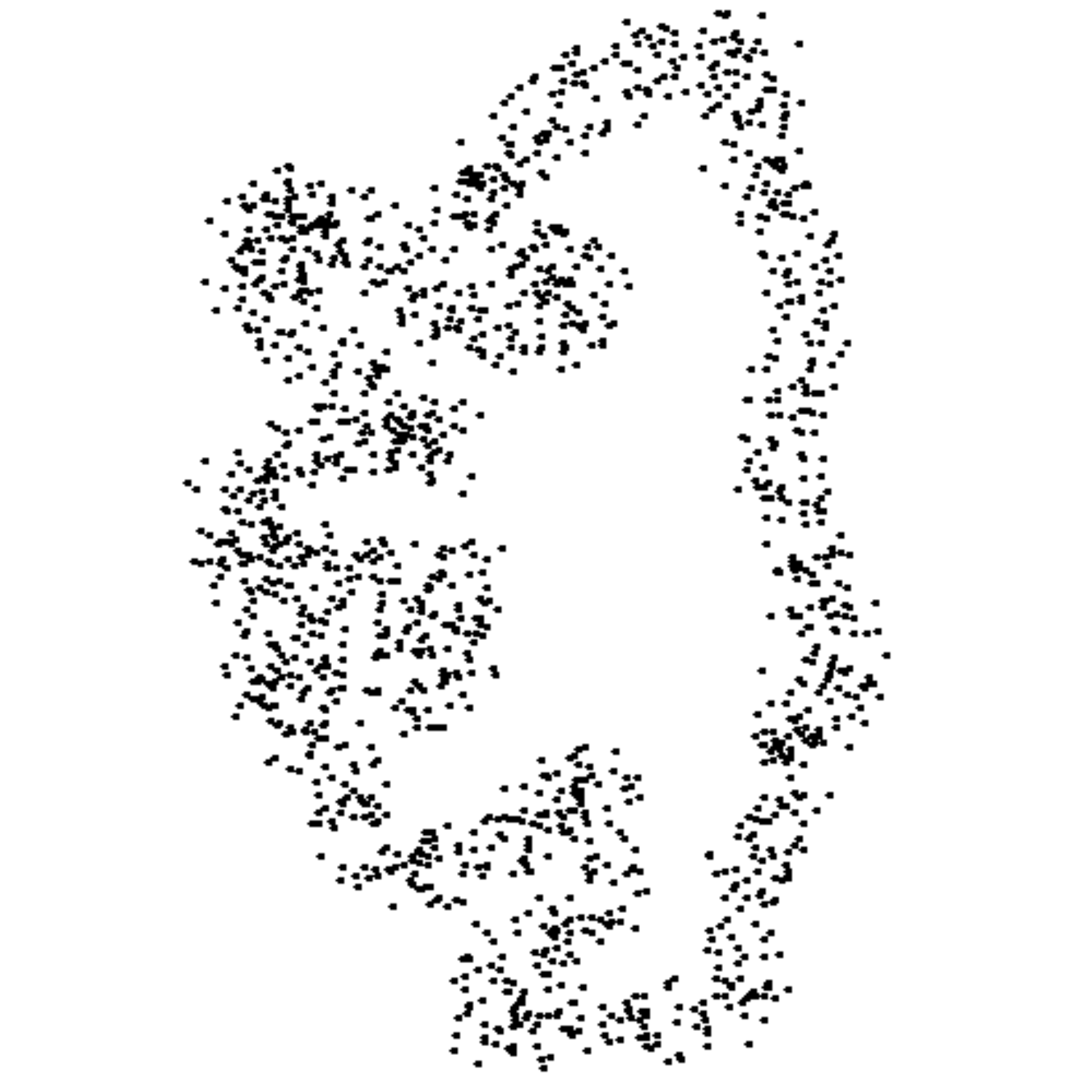}
\caption{$\quad P(\mbox{1 hole})\approx 66\%$, $\quad P(\mbox{2 holes})\approx 11\%$, 
 $P(\mbox{3 holes})\approx 3.8\%$, $\; P(\mbox{4 holes})\approx 3.3\%$,
 $\; P(\mbox{6 holes})\approx 1.1\%$. }
\label{fig:Mythical_Beast031-1662n6}
\end{figure}

\begin{figure}[h]
\includegraphics[height=0.5\linewidth]{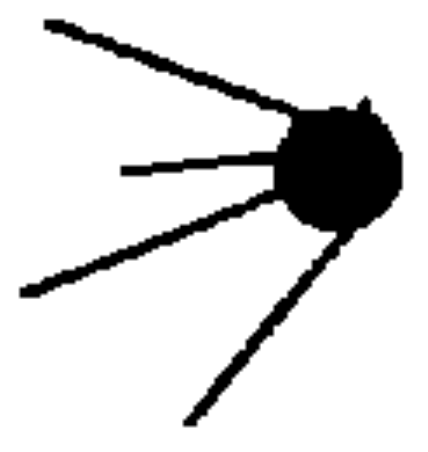}
\includegraphics[height=0.5\linewidth]{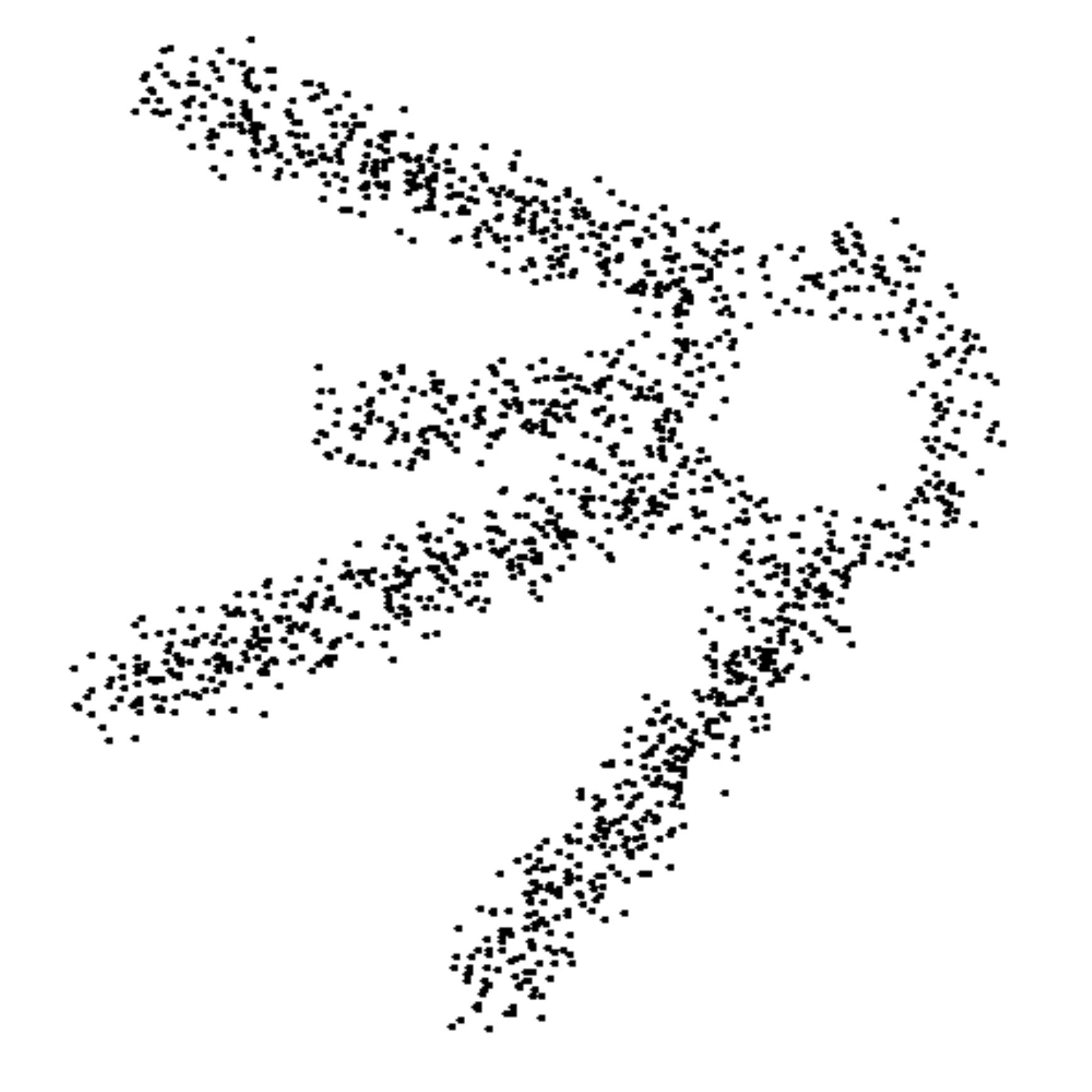}
\caption{$\quad P(\mbox{1 hole})\approx 58.3\%$, $\quad P(\mbox{2 holes})\approx 19.3\%$, 
 $P(\mbox{3 holes})\approx 4.2\%$, $\; P(\mbox{4 holes})\approx 1.6\%$,
 $\; P(\mbox{8 holes})\approx 0.8\%$. }
\label{fig:Satellite094-1687n5}
\end{figure}

\begin{figure}[h]
\includegraphics[height=0.6\linewidth]{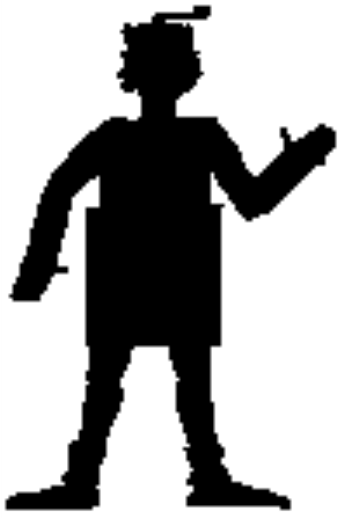}
\includegraphics[height=0.6\linewidth]{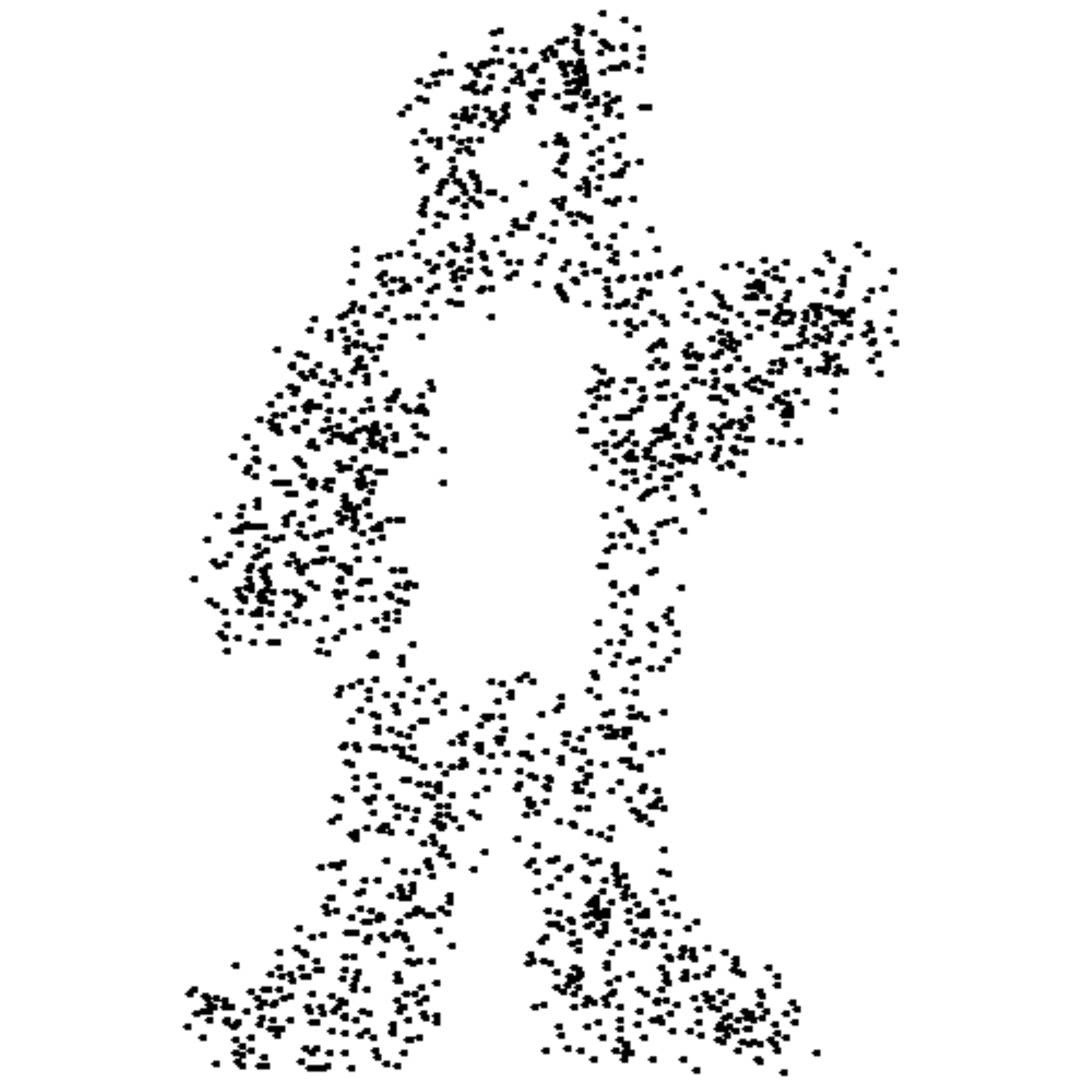}
\caption{$\quad P(\mbox{1 hole})\approx 49.6\%$, $\quad P(\mbox{2 holes})\approx 21.1\%$, 
 $P(\mbox{3 holes})\approx 4.7\%$, $\; P(\mbox{4 holes})\approx 3.3\%$,
 $\;P(\mbox{5 holes})\approx 1.8\%$. }
\label{fig:Male_Figure001-1823n6}
\end{figure}

\begin{figure}[h]
\includegraphics[width=0.5\linewidth]{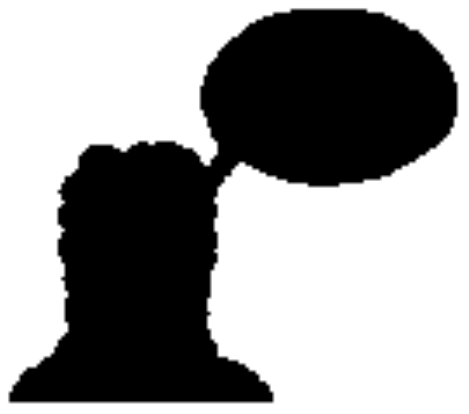}
\includegraphics[width=0.46\linewidth]{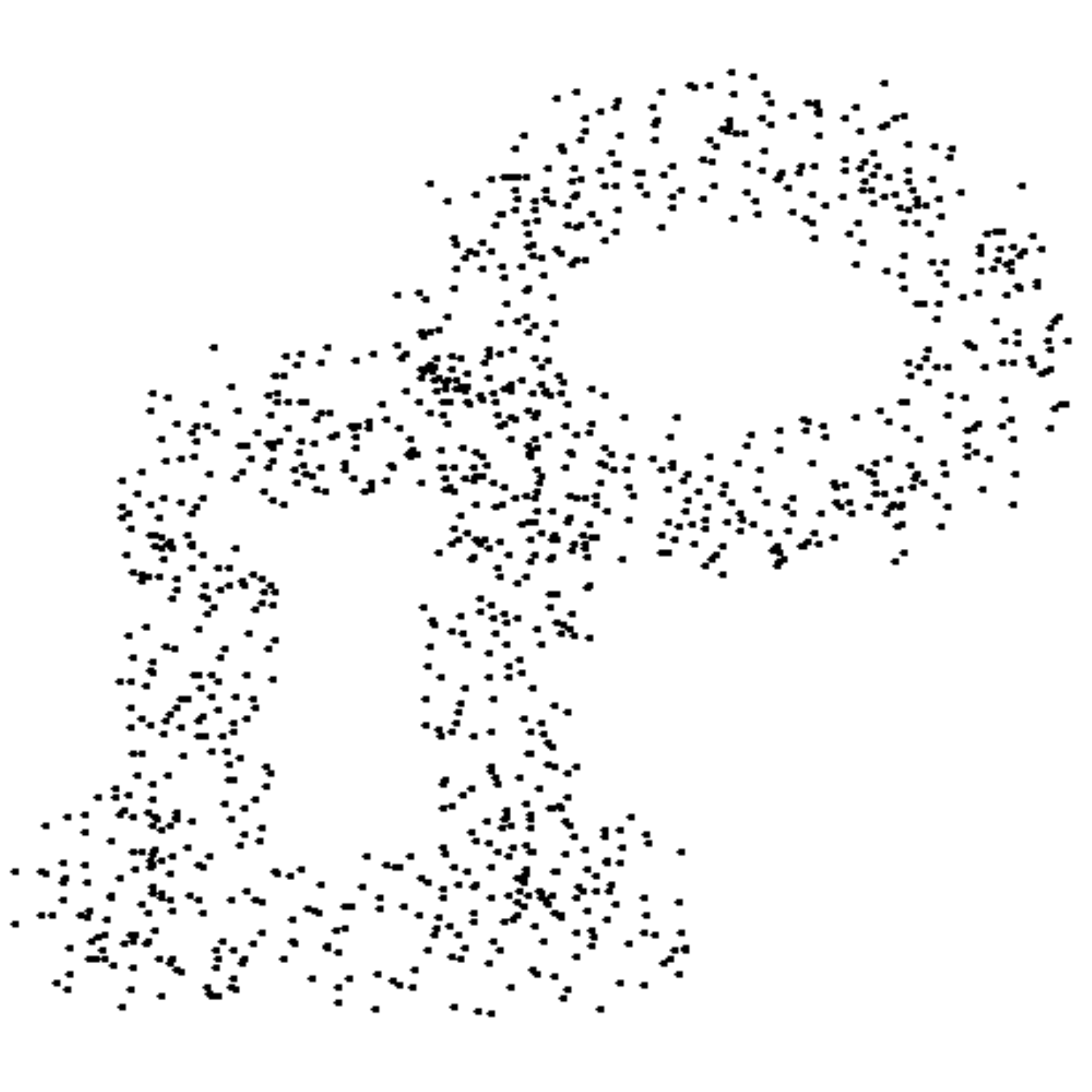}
\caption{$\quad P(\mbox{2 holes})\approx 43.7\%$, $\quad P(\mbox{1 hole})\approx 27.8\%$, 
 $P(\mbox{3 holes})\approx 2.5\%$, $\; P(\mbox{5 holes})\approx 2.1\%$,
 $\;P(\mbox{6 holes})\approx 1.6\%$. }
\label{fig:Thought_Bubble001}
\end{figure}

The left hand side pictures in 
 Fig.~\ref{fig:Tree004-1217n10}--\ref{fig:Thought_Bubble001}
 are from \url{http://www.lems.brown.edu/~dmc}.
The right hand side pictures are extracted contours with added noise.
 
\begin{figure*}[h]
\includegraphics[width=0.32\linewidth]{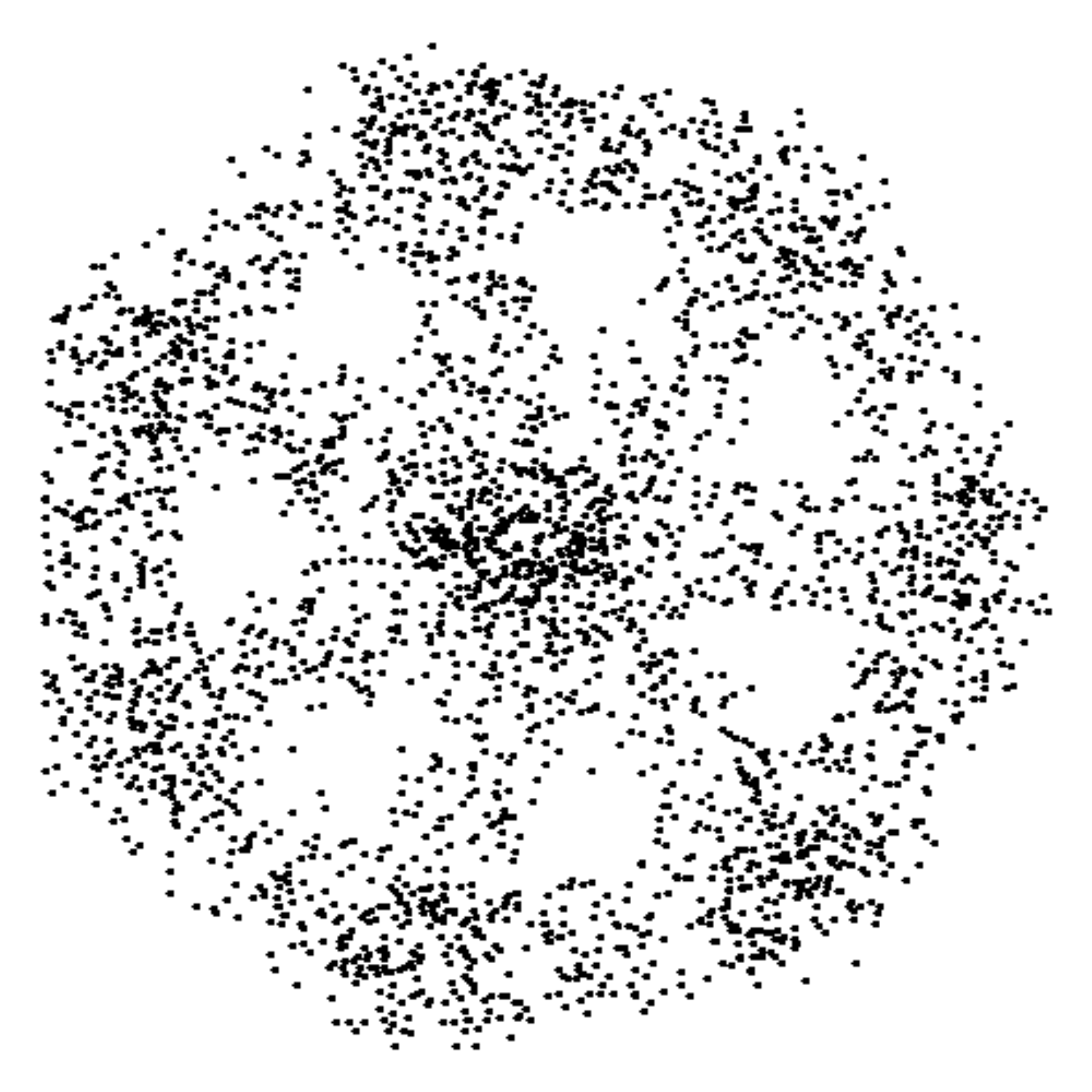}
\hspace*{0.02\linewidth}
\includegraphics[width=0.32\linewidth]{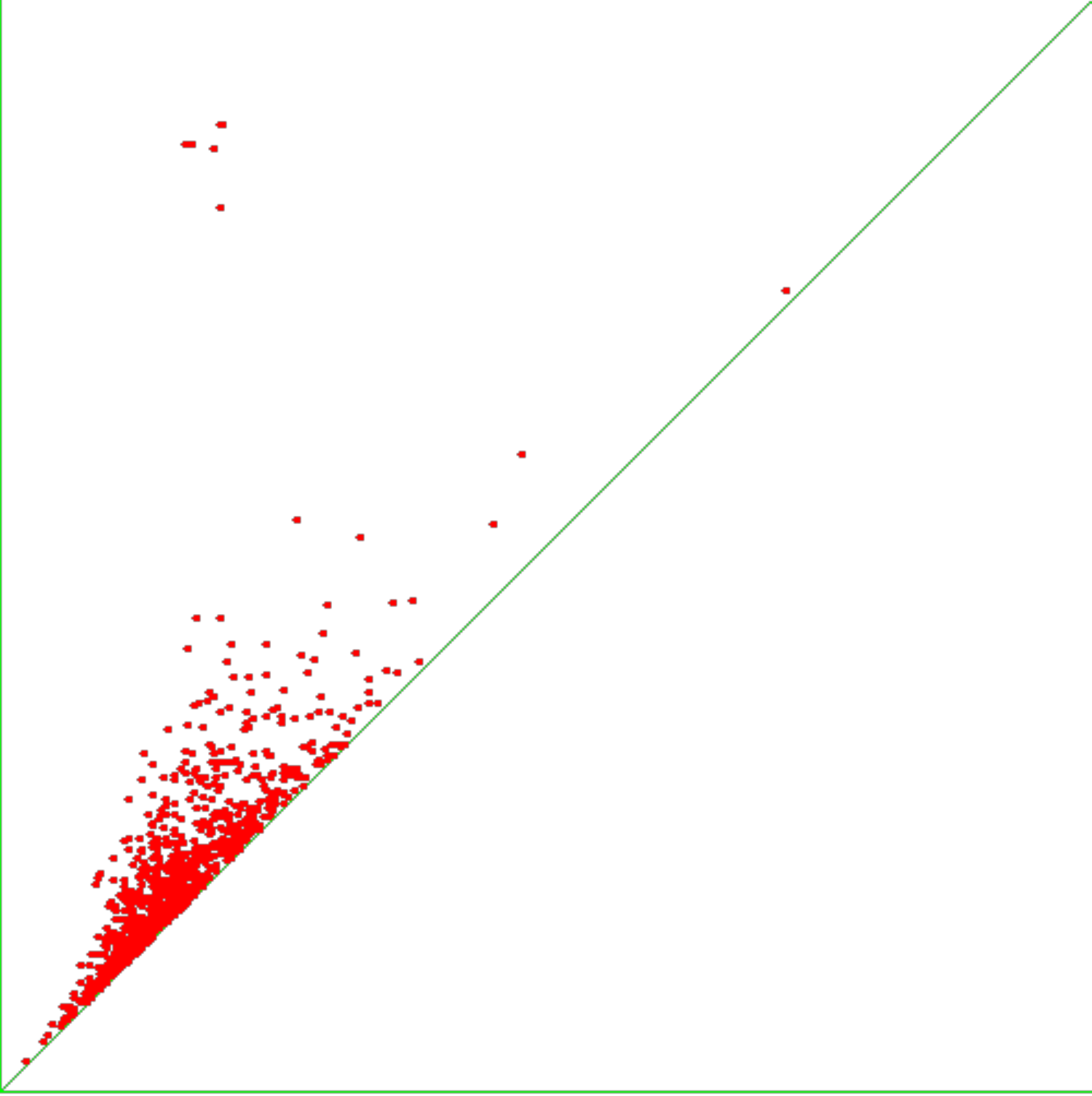}
\hspace*{0.02\linewidth}
\includegraphics[width=0.32\linewidth]{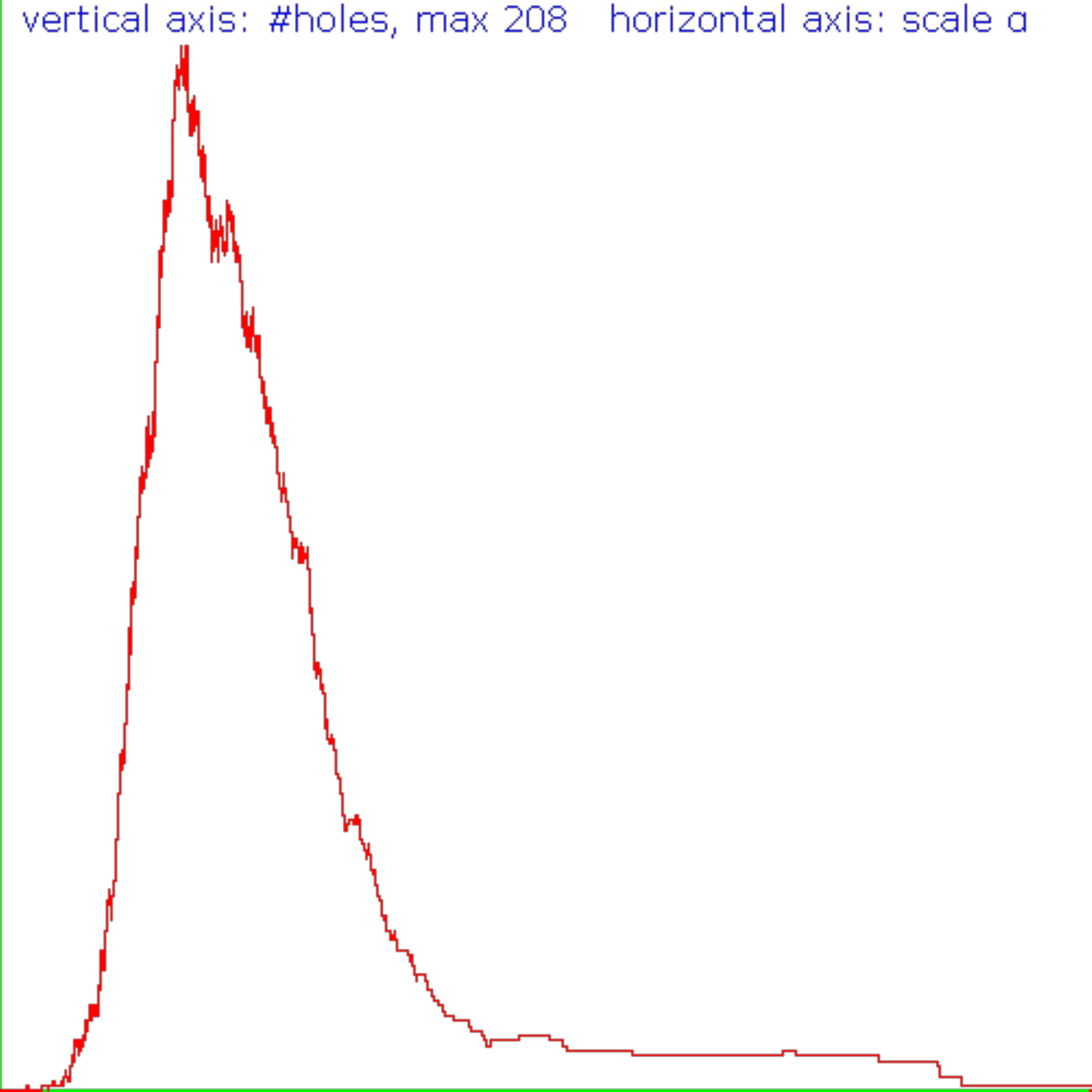}
\caption{$\quad P(\mbox{7 holes})\approx 22\%$, $\quad P(\mbox{1 hole})\approx 14\%$, 
 $\quad P(\mbox{8 holes})\approx 7.5\%$, $\quad P(\mbox{6 holes})\approx 5.8\%$,
 $\quad P(\mbox{10 holes})\approx 4.4\%$. }
\label{fig:wheel7N3265n62}
\end{figure*}

\begin{figure*}[h]
\includegraphics[width=0.32\linewidth]{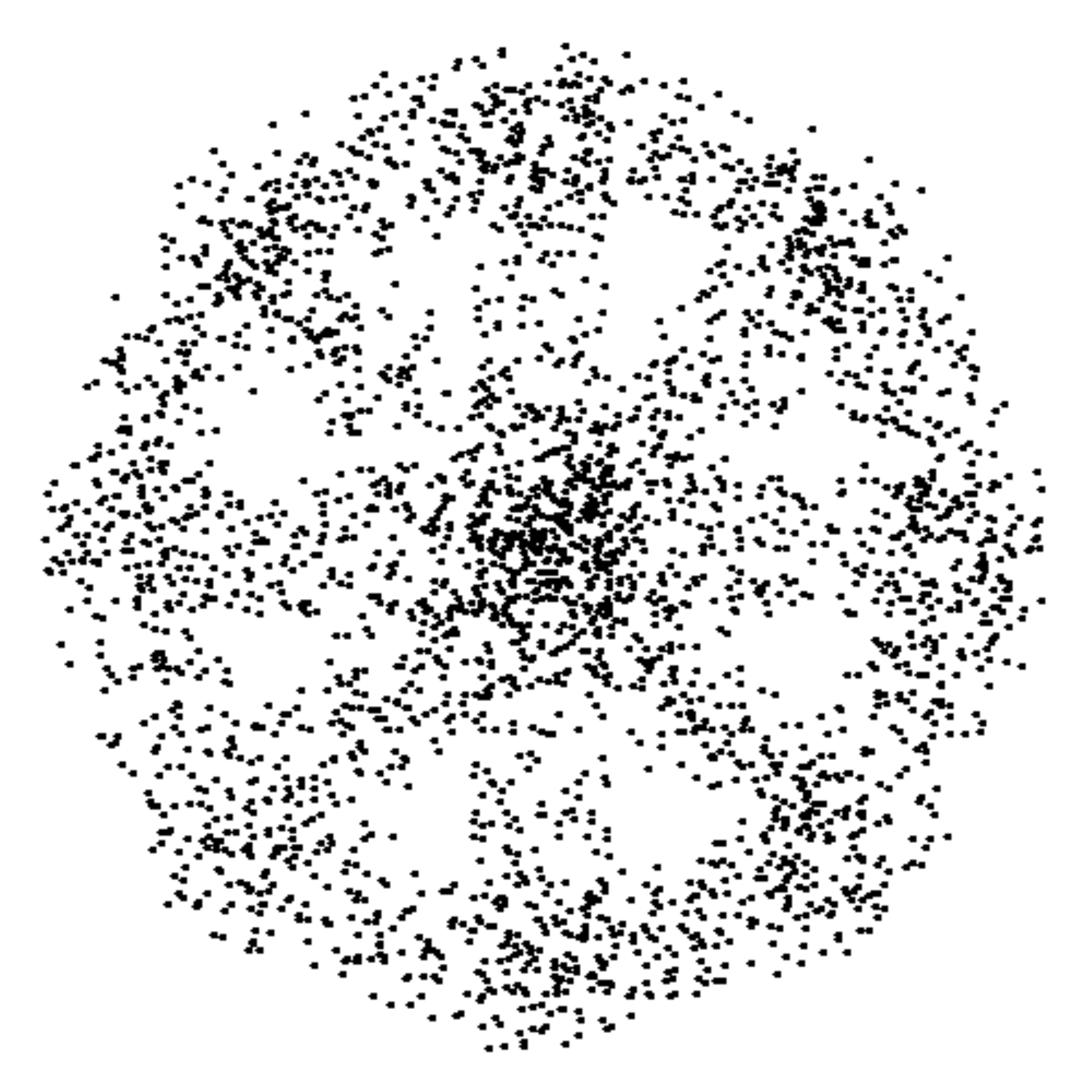}
\hspace*{0.02\linewidth}
\includegraphics[width=0.32\linewidth]{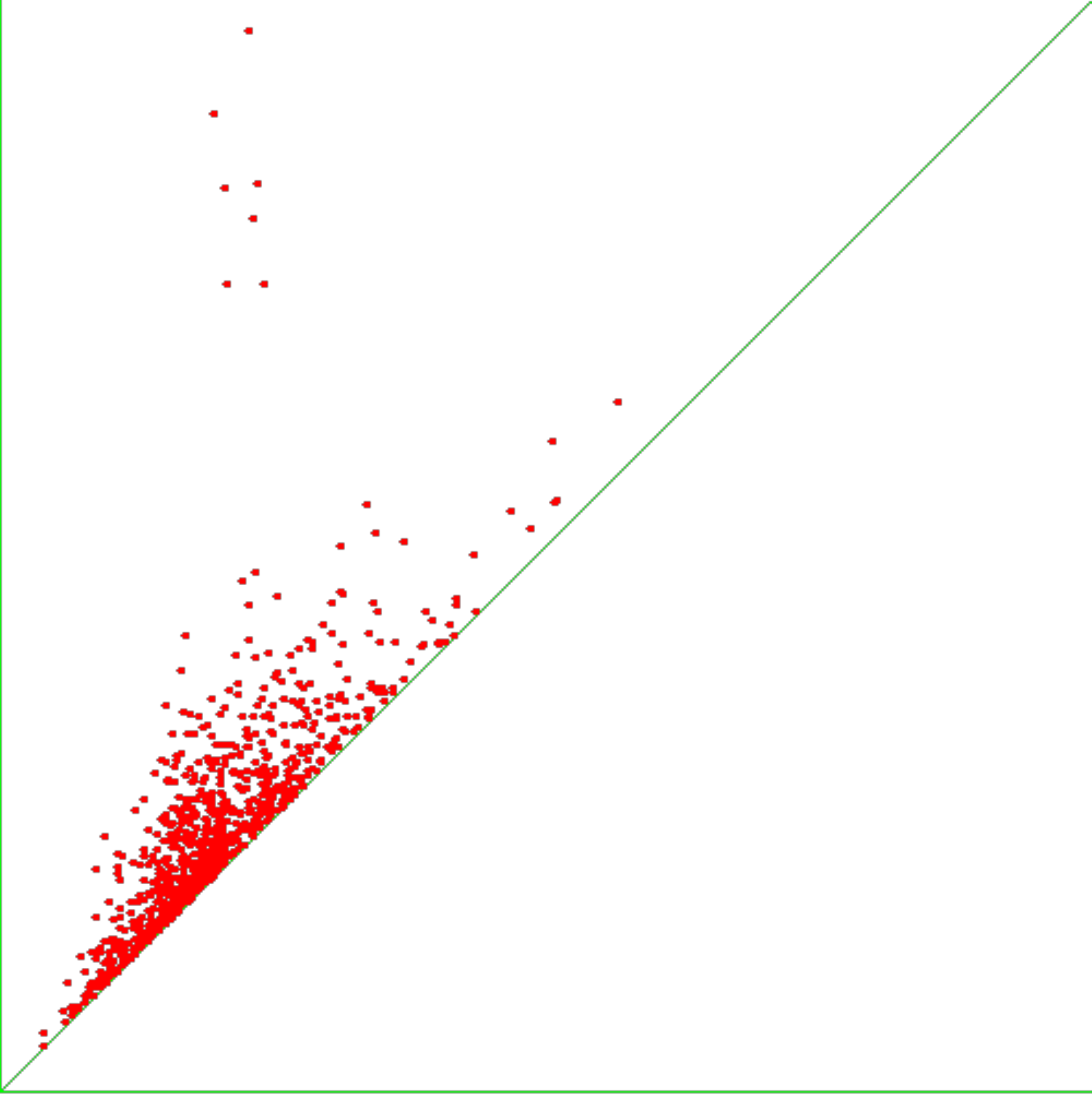}
\hspace*{0.02\linewidth}
\includegraphics[width=0.32\linewidth]{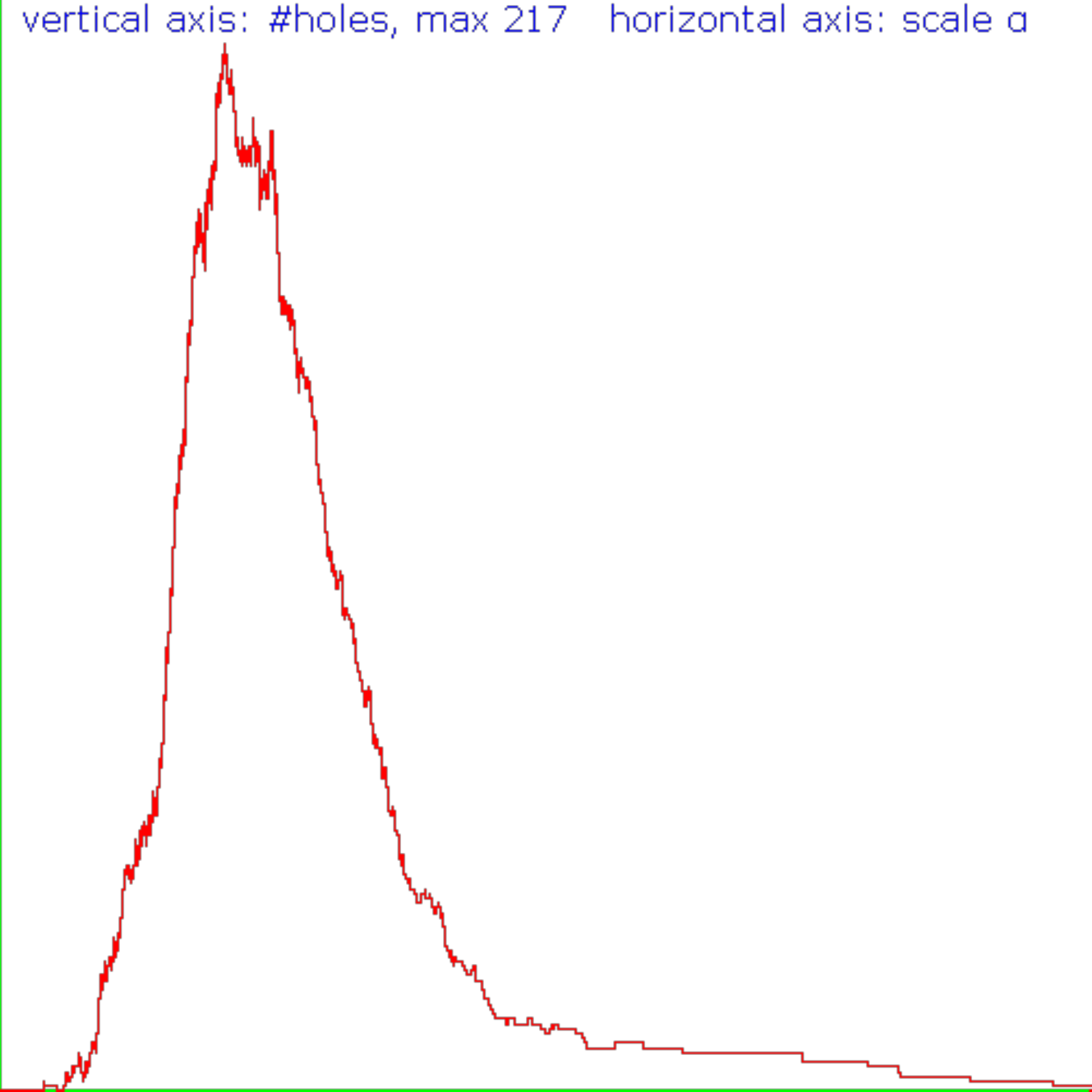}
\caption{$\quad P(\mbox{8 holes})\approx 11.5\%$, $\quad P(\mbox{2 holes})\approx 8.5\%$, 
 $\quad P(\mbox{3 holes})\approx 7\%$, $\quad P(\mbox{9 holes})\approx 6.8\%$,
 $\quad P(\mbox{6 holes})\approx 6.5\%$. }
\label{fig:wheel8N3265n63}
\end{figure*}

\begin{figure*}[h]
\includegraphics[width=0.32\linewidth]{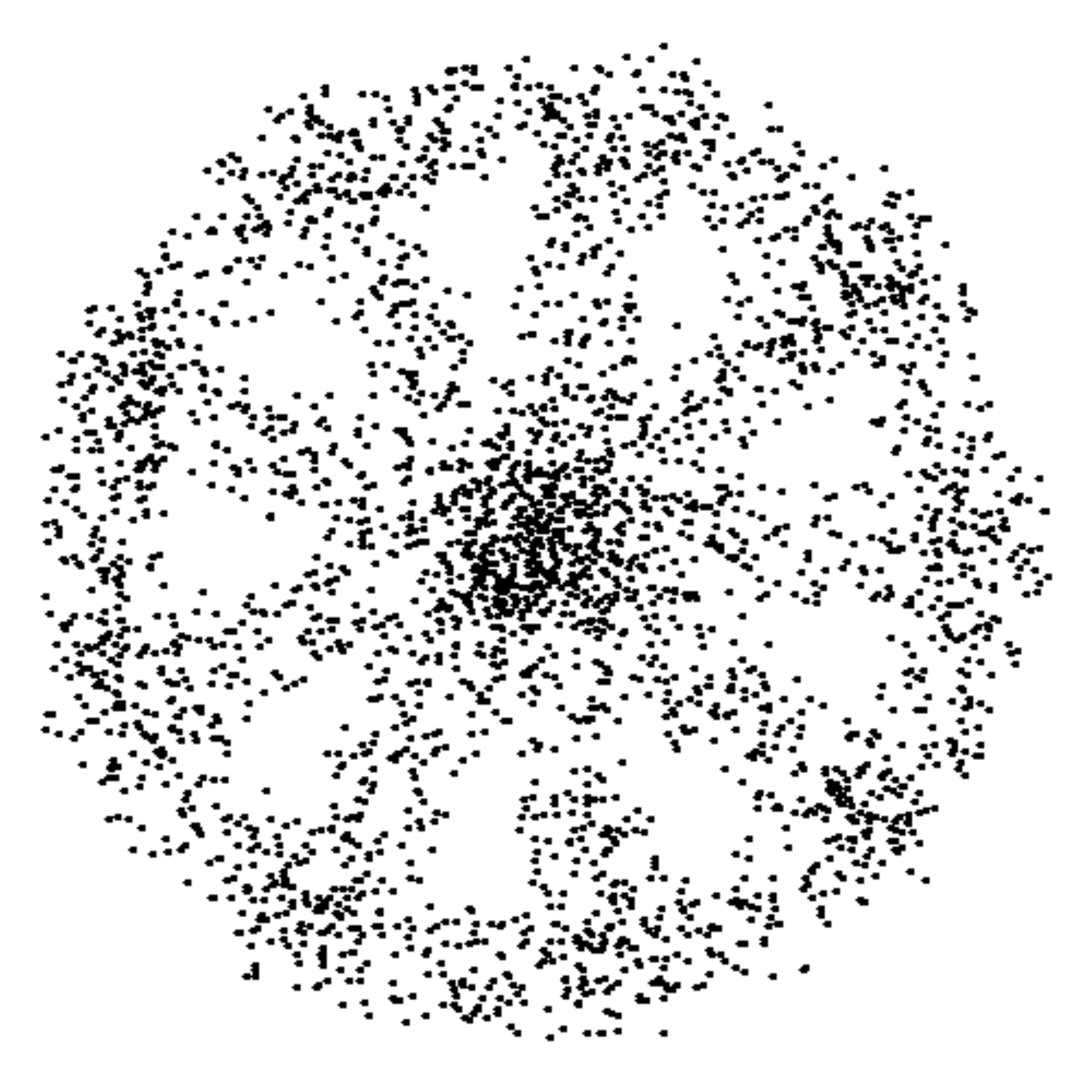}
\hspace*{0.02\linewidth}
\includegraphics[width=0.32\linewidth]{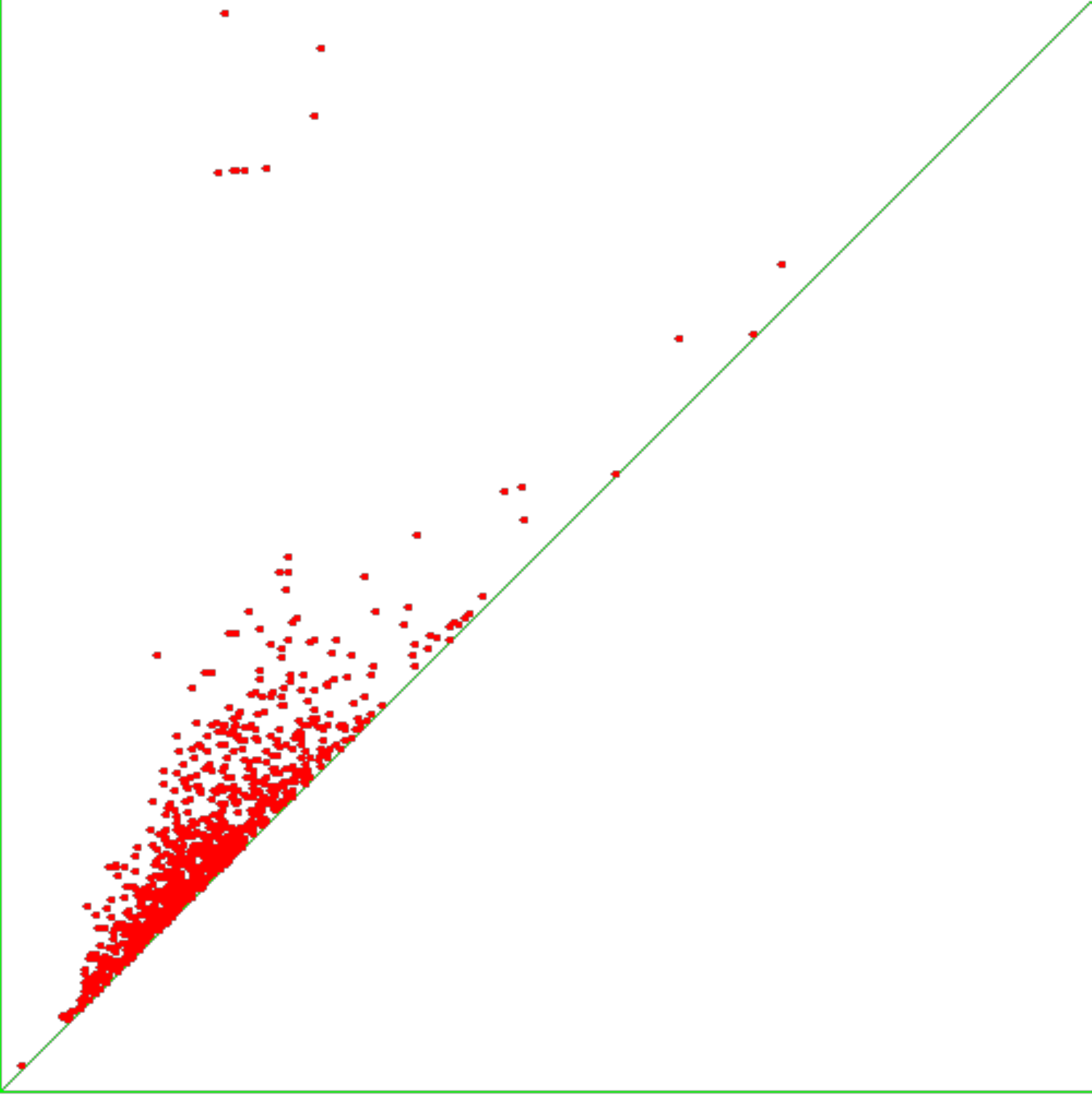}
\hspace*{0.02\linewidth}
\includegraphics[width=0.32\linewidth]{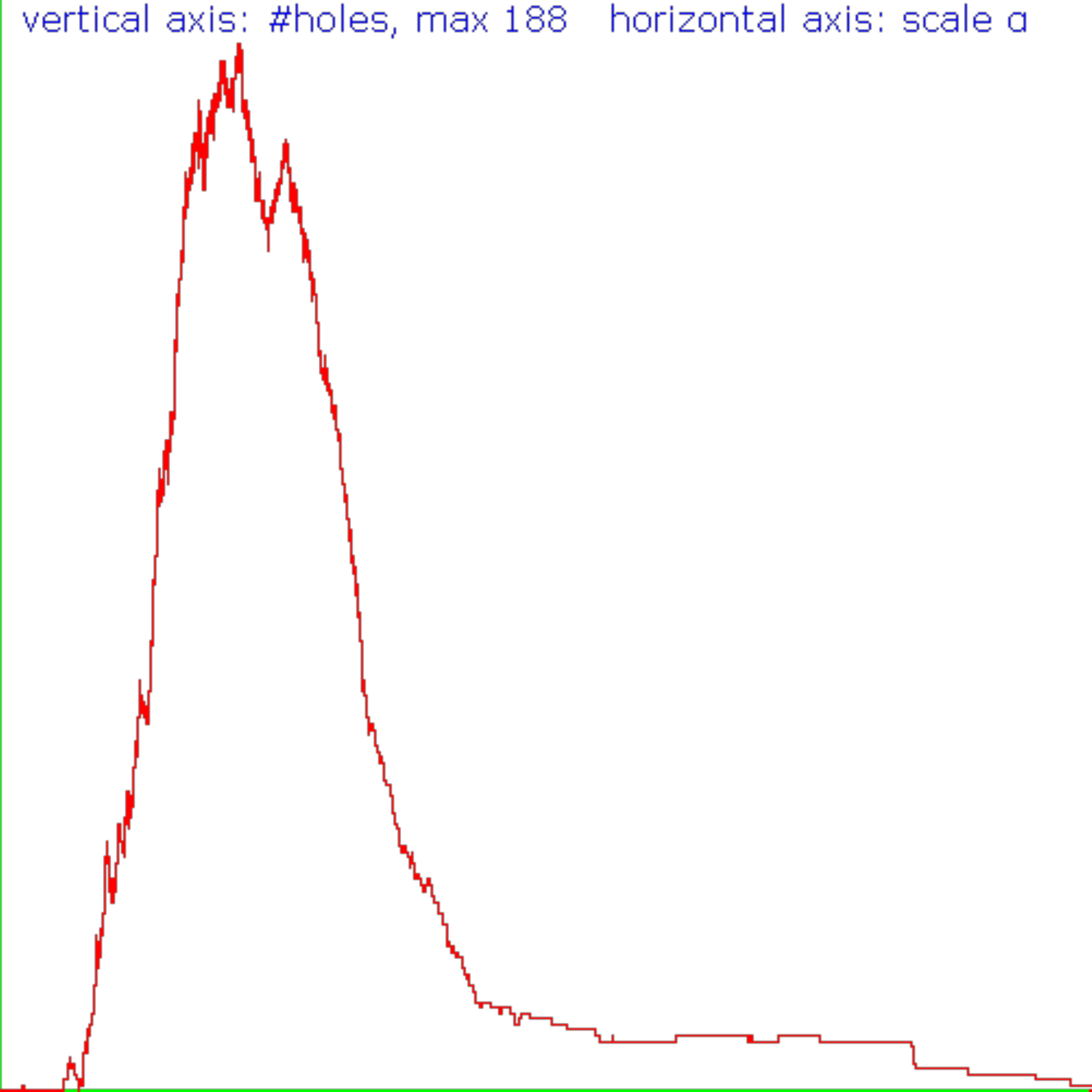}
\caption{$\quad P(\mbox{9 holes})\approx 18.5\%$, $\quad P(\mbox{10 holes})\approx 11.3\%$, 
 $\quad P(\mbox{3 holes})\approx 6.8\%$, $\quad P(\mbox{3 holes})\approx 6.8\%$,
 $\quad P(\mbox{4 holes})\approx 5.3\%$. }
\label{fig:wheel9N3265n54}
\end{figure*}

The left hand side pictures in Fig.~\ref{fig:wheel7N3265n62}--\ref{fig:wheel9N3265n54}
 contain a cloud $C$ uniformly generated around \emph{wheels} 
 (the boundaries of regular polygons with the radii to all vertices).
The middle pictures show the persistence diagrams $\PD\{C^{\al}\}$.
The right hand side pictures are the staircases $\PS\{C^{\al}\}$
 giving the number of holes of $C$ depending on the scale $\al$. 
\medskip
 
\begin{figure*}[h]
\includegraphics[width=0.32\linewidth]{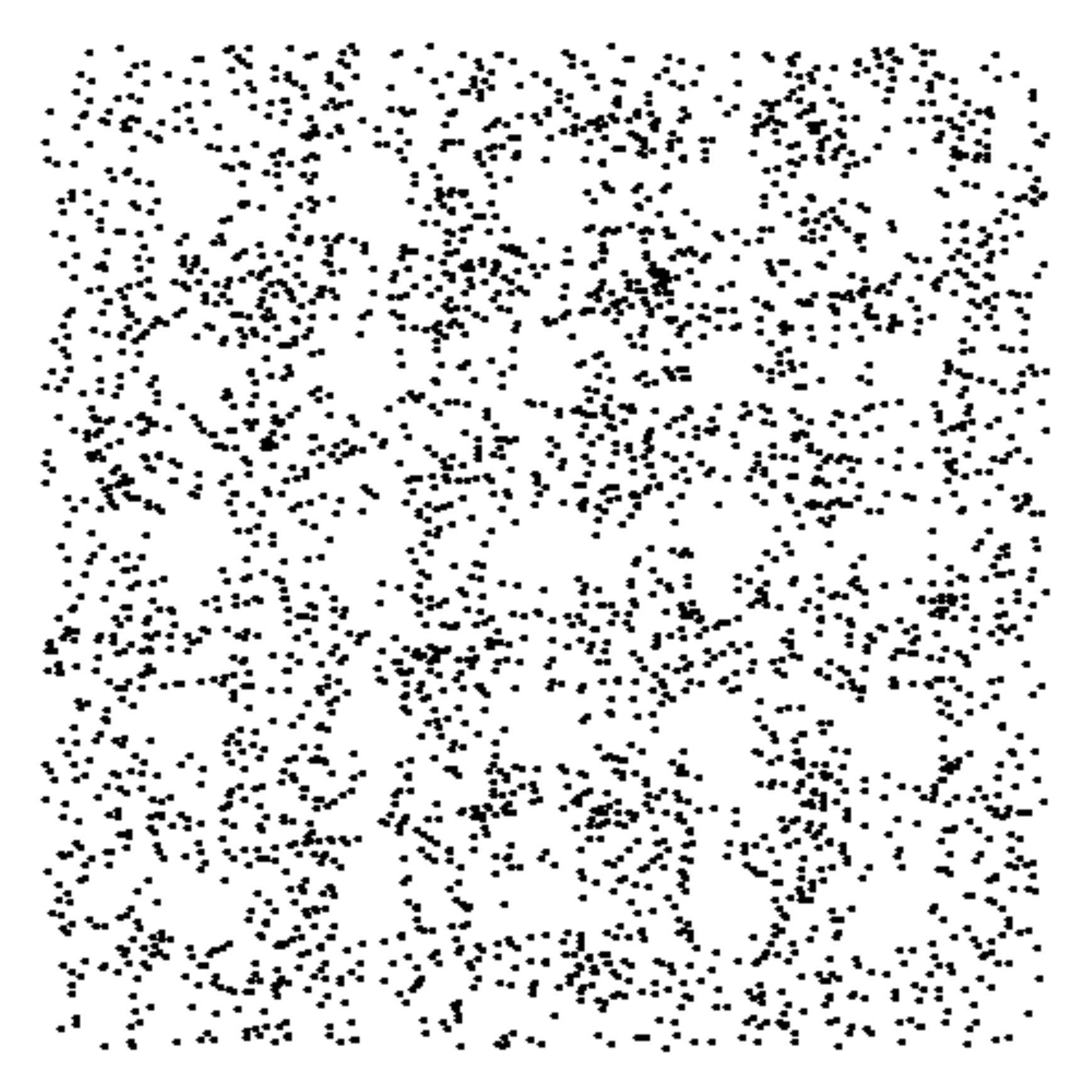}
\hspace*{0.02\linewidth}
\includegraphics[width=0.32\linewidth]{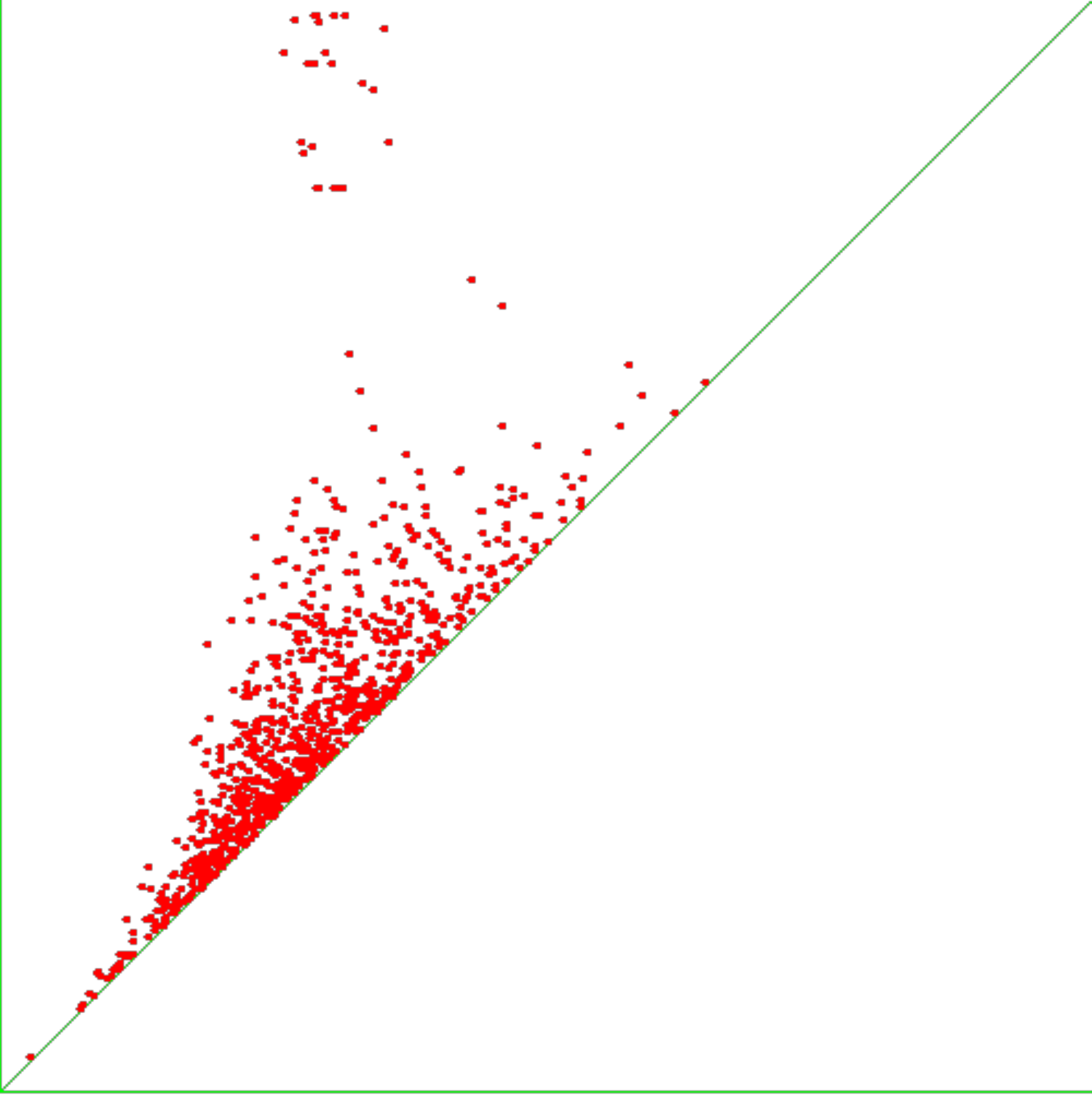}
\hspace*{0.02\linewidth}
\includegraphics[width=0.32\linewidth]{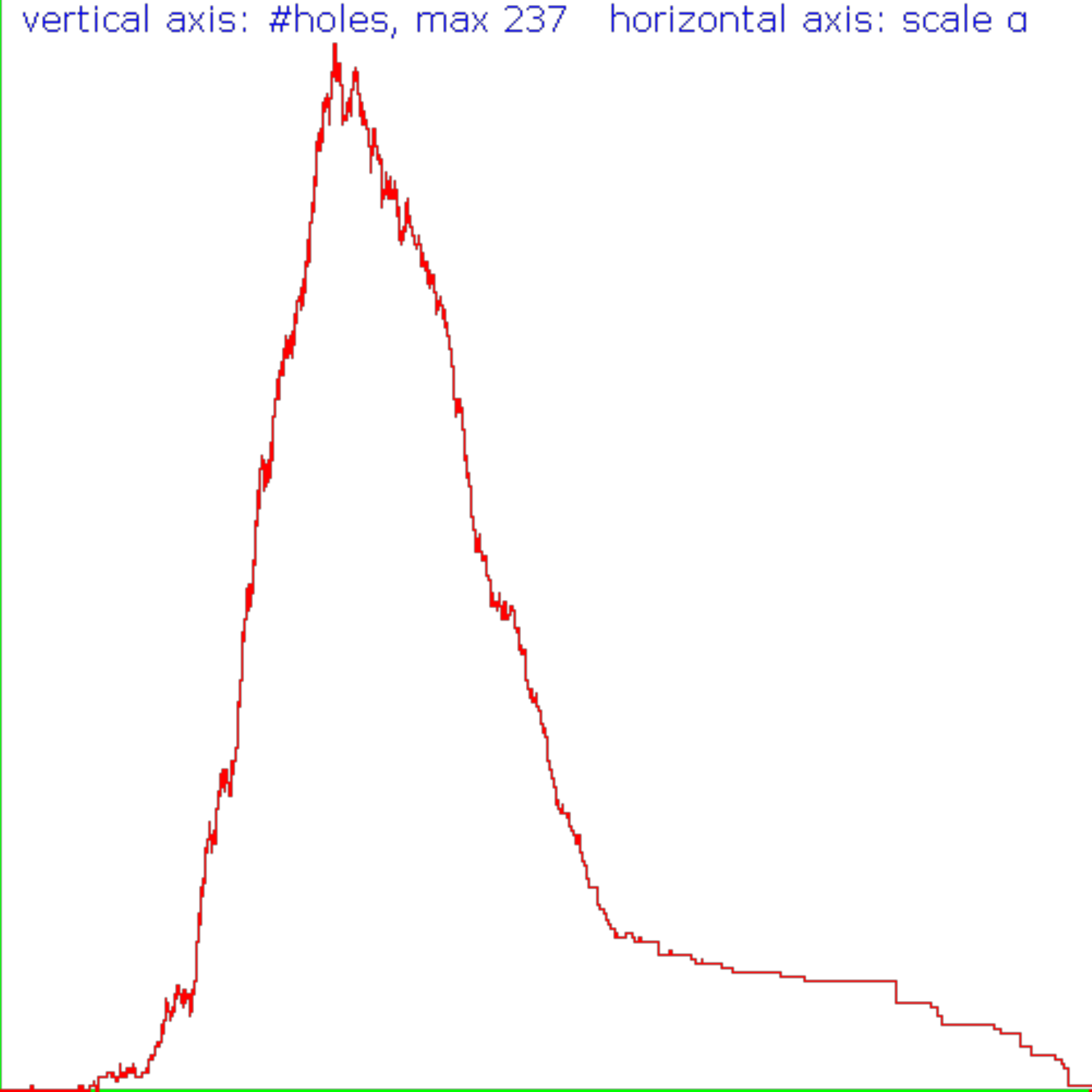}
\caption{$\quad P(\mbox{25 holes})\approx 8.8\%$, $\quad P(\mbox{0 holes})\approx 5.4\%$, 
 $\quad P(\mbox{15 holes})\approx 5\%$, $\quad P(\mbox{27 holes})\approx 4.6\%$,
 $\quad P(\mbox{20 holes})\approx 3.5\%$. }
\label{fig:lattice7N3265n52}
\end{figure*}

\begin{figure*}[h]
\includegraphics[width=0.32\linewidth]{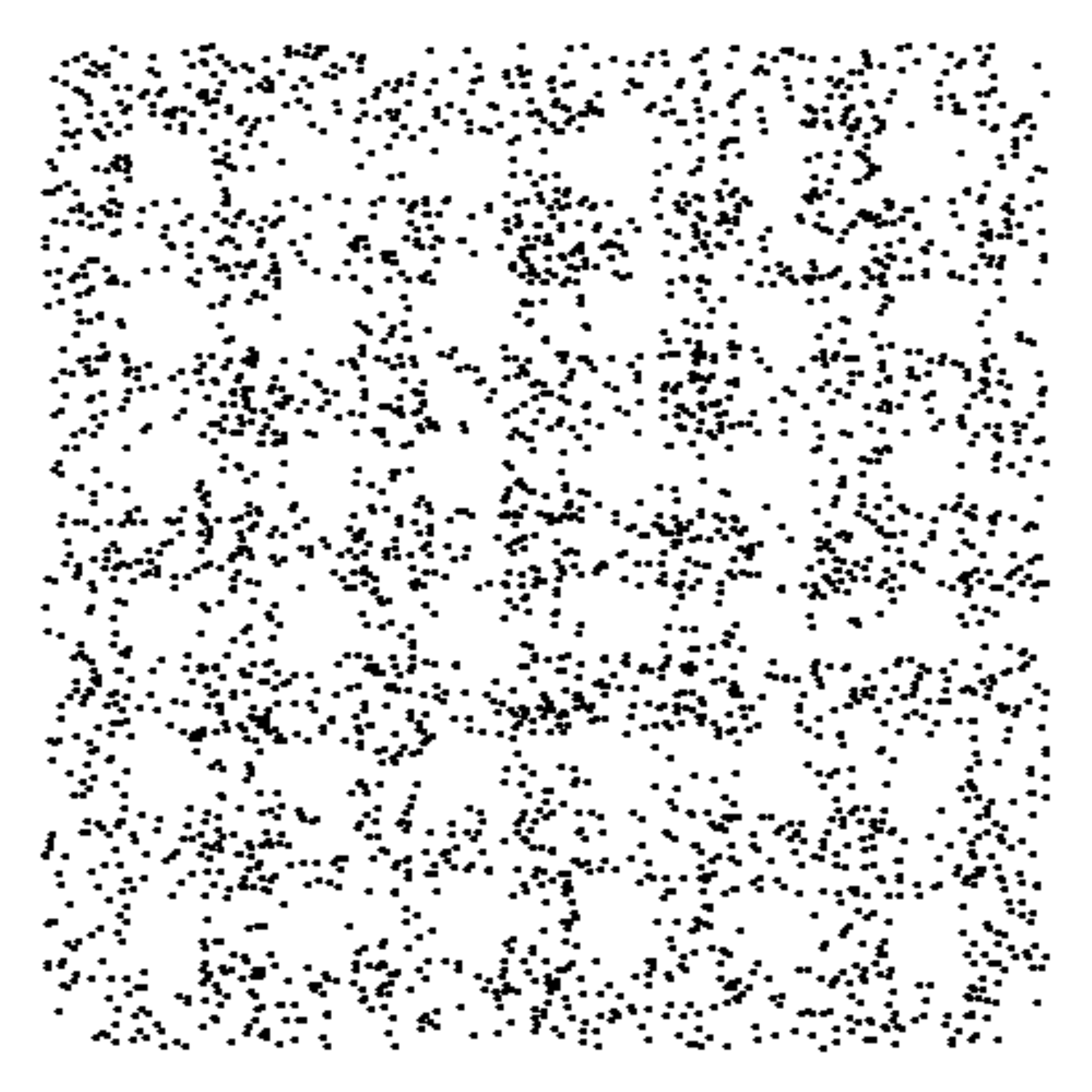}
\hspace*{0.02\linewidth}
\includegraphics[width=0.32\linewidth]{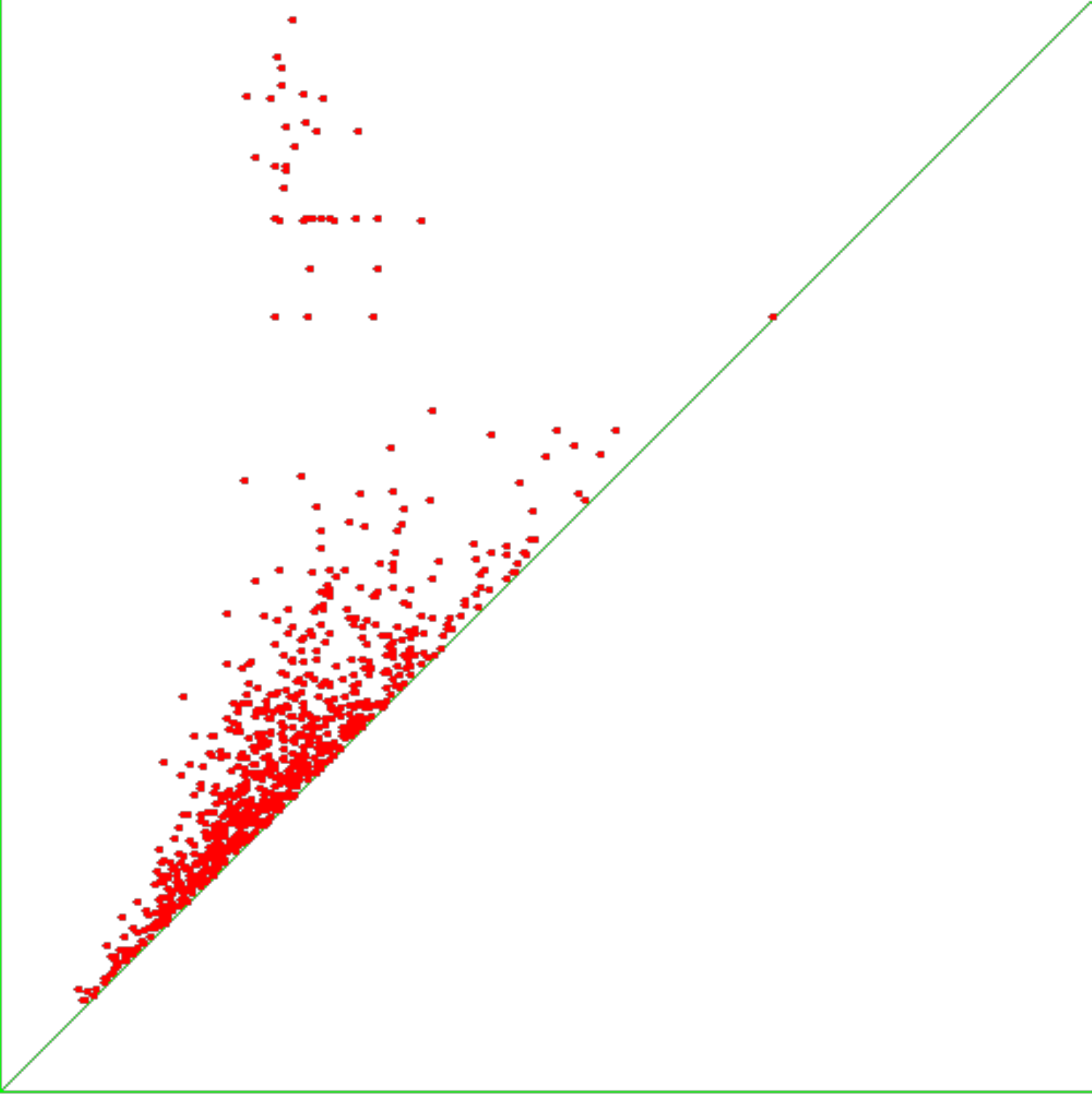}
\hspace*{0.02\linewidth}
\includegraphics[width=0.32\linewidth]{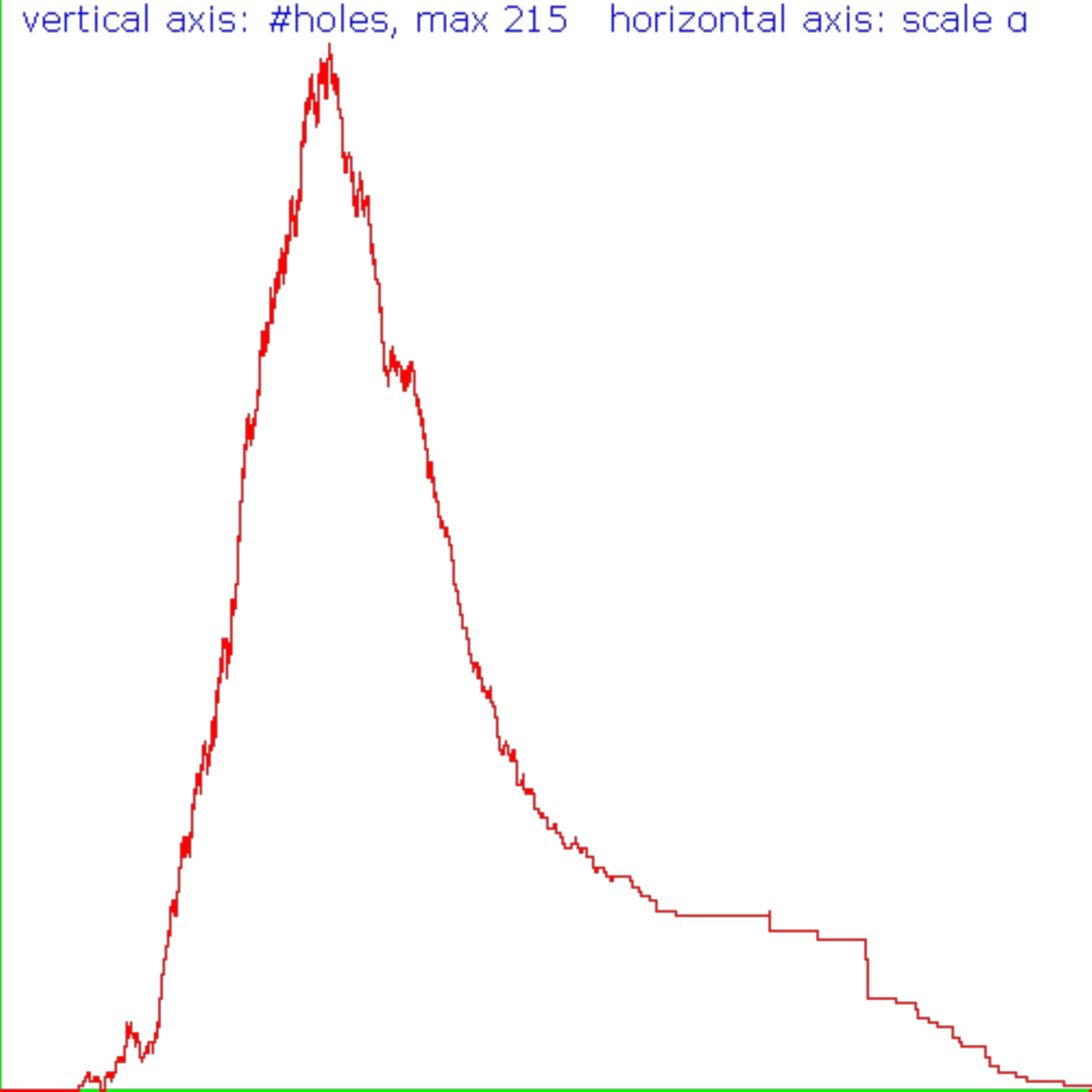}
\caption{$\quad P(\mbox{36 holes})\approx 9.4\%$, $\quad P(\mbox{31 holes})\approx 4.8\%$, 
 $\quad P(\mbox{33 holes})\approx 4.8\%$, $\quad P(\mbox{2 holes})\approx 4.6\%$,
 $\quad P(\mbox{1 hole})\approx 3.2\%$. }
\label{fig:lattice8N3265n39}
\end{figure*}

\begin{figure*}[h]
\includegraphics[width=0.32\linewidth]{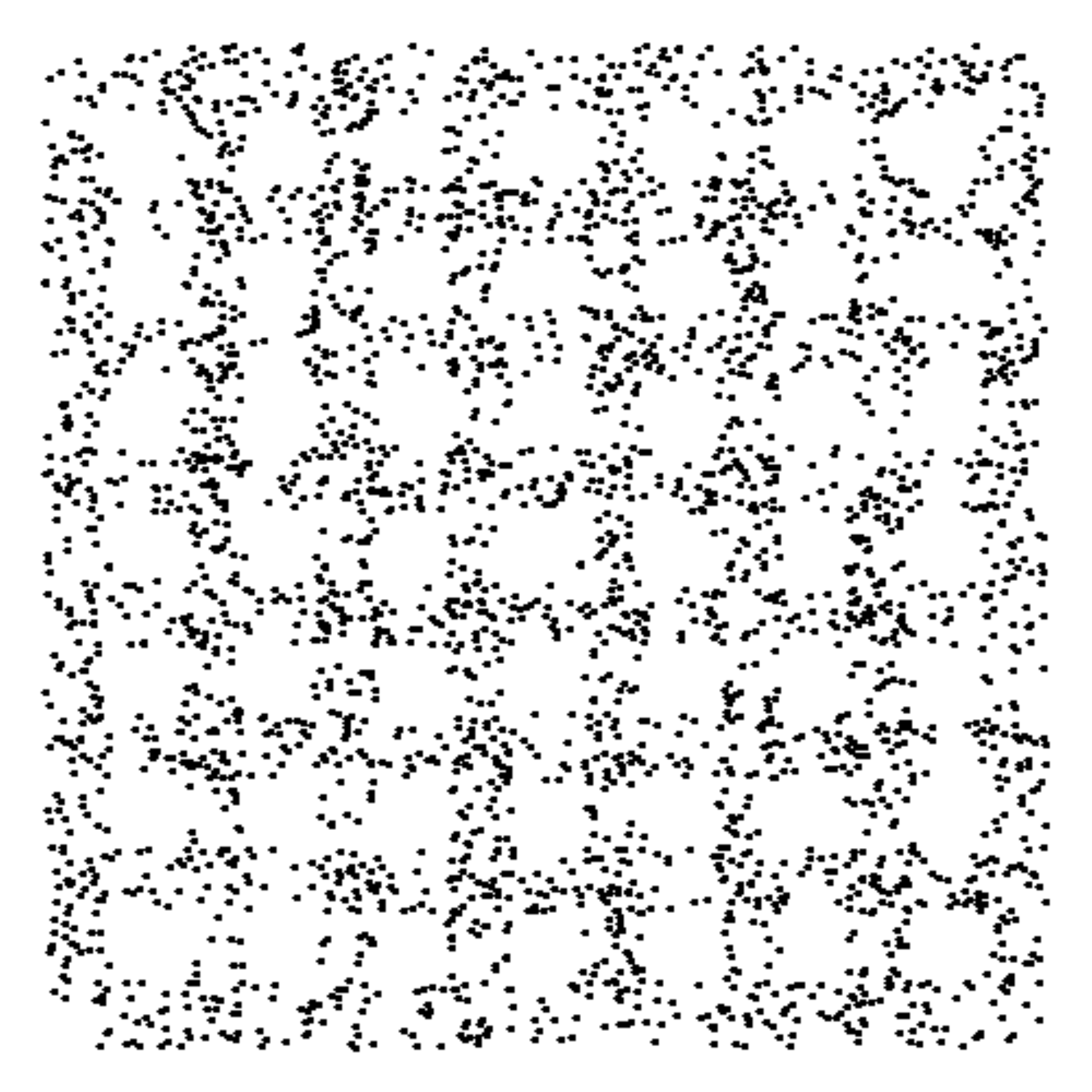}
\hspace*{0.02\linewidth}
\includegraphics[width=0.32\linewidth]{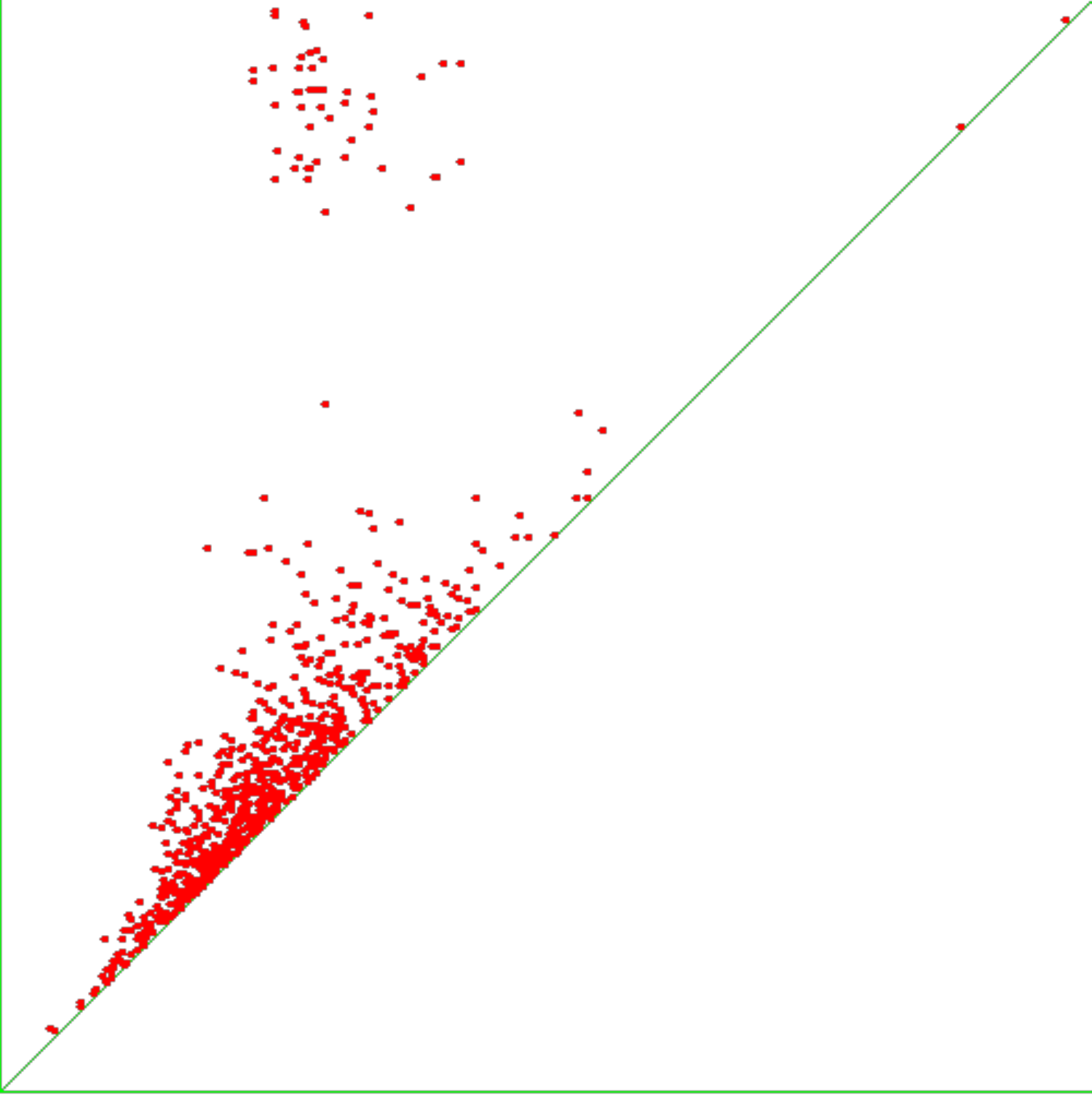}
\hspace*{0.02\linewidth}
\includegraphics[width=0.32\linewidth]{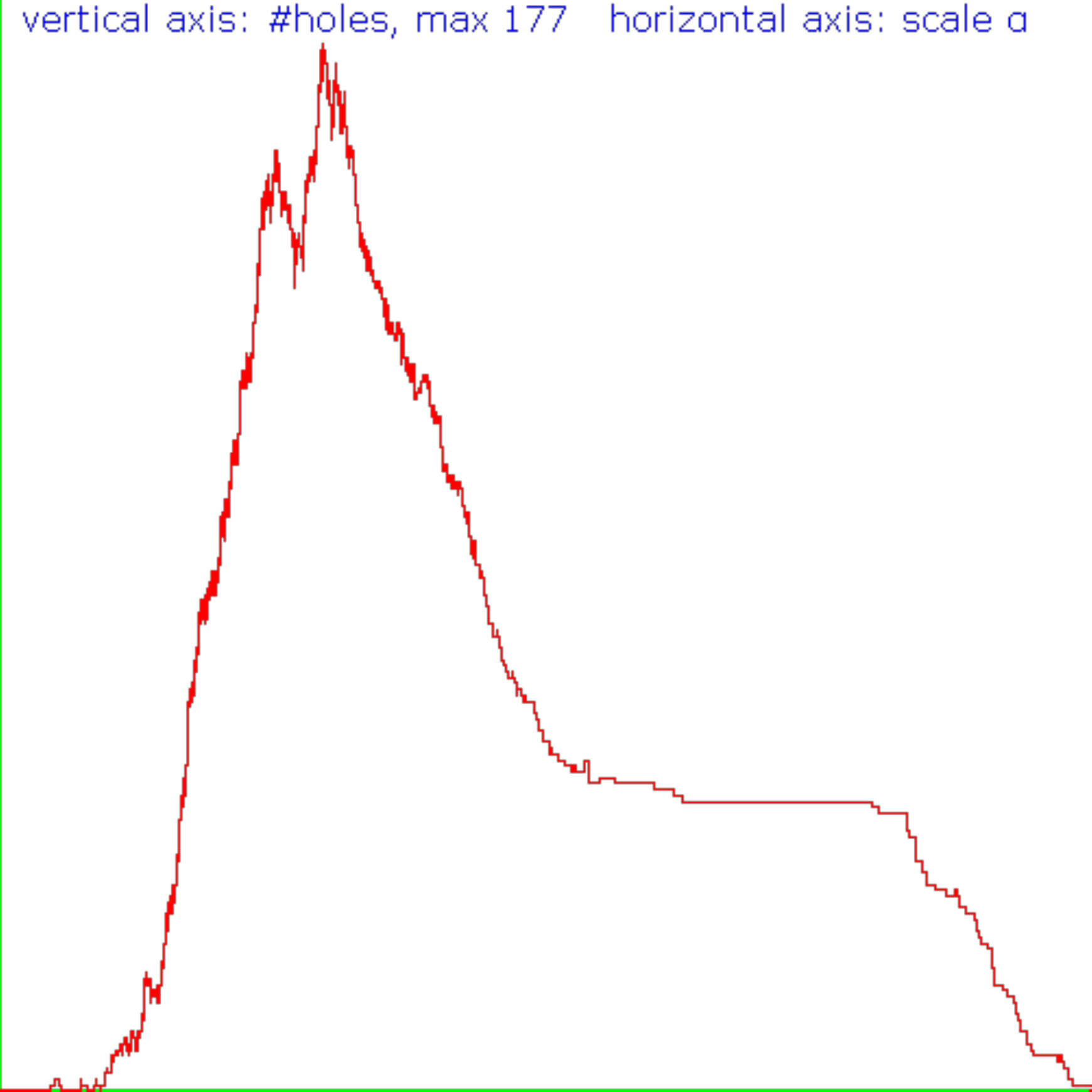}
\caption{$\quad P(\mbox{49 holes})\approx 18.4\%$, $\quad P(\mbox{52 holes})\approx 4.9\%$, 
 $\quad P(\mbox{1 hole})\approx 3.3\%$, $\quad P(\mbox{6 holes})\approx 2.8\%$,
 $\quad P(\mbox{0 holes})\approx 2.7\%$. }
\label{fig:lattice9N3265n28}
\end{figure*}

The left hand side pictures in Fig.~\ref{fig:lattice7N3265n52}--\ref{fig:lattice9N3265n28}
 are noisy clouds around square lattices containing $25,36,49$ small squares.
The algorithm \alg finds the expected number $49$ of holes in
 Fig.~\ref{fig:lattice9N3265n28} when even humans may struggle.

\end{document}